\documentclass{aa}
\usepackage{natbib}
\bibpunct{(}{)}{;}{a}{}{,}
\usepackage{graphicx}
\usepackage[varg]{txfonts}
\usepackage{soul}
\usepackage{booktabs}
\usepackage{etoolbox}
\usepackage{orcidlink}
\usepackage{physics}
\usepackage{comment}
\usepackage{hyperref}
\usepackage[all]{hypcap}
\usepackage[T1]{fontenc}
\usepackage{xcolor}
\usepackage{subfigure}
\usepackage[normalem]{ulem}
\usepackage{cancel}
\usepackage{multirow}
\usepackage{amsmath}
\usepackage{xcolor}
\usepackage{float}

\usepackage[switch]{lineno}

\makeatletter
\renewcommand*\aa@pageof{, page \thepage{} of \pageref*{LastPage}}
\makeatother

\hypersetup{
    hidelinks,
    colorlinks=true,
    linkcolor=blue,
    urlcolor=blue,
    citecolor=blue,
}

\pdfoutput=1
\newcommand{\vide}{\tt VIDE\normalfont\xspace}

\begin{document}

   \title{Extracting the Alcock-Paczy\'nski signal from voids:\\ A novel approach via reconstruction}
   
   \author{G. Degni
   \inst{1,2}\thanks{\email{degni@cppm.in2p3.fr}}\orcidlink{0009-0001-4912-1087}
          \and
          E. Sarpa\inst{3,4,5}\fnmsep \orcidlink{https://orcid.org/0000-0002-1256-655X}
          \and 
          M. Aubert\inst{6}\fnmsep \orcidlink{https://orcid.org/0009-0002-7667-8814}
          \and 
          E. Branchini\inst{7,8,9}\fnmsep \orcidlink{https://orcid.org/0000-0002-0808-6908}
          \and 
          A. Pisani\inst{1,10}\fnmsep \orcidlink{https://orcid.org/0000-0002-6146-4437}
          \and
          H.M. Courtois\inst{11}\fnmsep \orcidlink{https://orcid.org/0000-0003-0509-1776}
          }

   \institute{Aix-Marseille Universit\'e, CNRS/IN2P3, CPPM, Marseille, France
   \and 
   Dipartimento di Fisica, Università di Roma Tre, Via della Vasca Navale 84, I-00146 Roma,
Italy
             \and
             SISSA, International School for Advanced Studies, Via Bonomea 265, 34136 Trieste TS, Italy
             \and
             ICSC - Centro Nazionale di Ricerca in High Performance Computing, Big Data e Quantum Computing, Via Magnanelli 2, Bologna, Italy
             \and
             INFN, Sezione di Trieste, Via Valerio 2, 34127 Trieste TS, Italy
             \and
             Universit\'e Clermont Auvergne, CNRS/IN2P3, LPCA, F-63000 Clermont-Ferrand, France
             \and
             Department of Physics, Università di Genova, Via Dodecaneso 33, 16146 Genova, Italy
             \and
             Istituto Nazionale di Fisica Nucleare, Sezione di Genova, Via Dodecaneso 33, 16146 Genova, Italy
             \and
            INAF-Osservatorio Astronomico di Brera. Via Brera 28, 20122, Milano, Italy
             \and
            Department of Astrophysical Sciences, Peyton Hall, Princeton University, Princeton, NJ 08544, USA
             \and
             Universit\'e Claude Bernard Lyon 1, IUF, IP2I Lyon, 4 rue Enrico Fermi, 69622 Villeurbanne, France
             }

   \date{Received 18 September 2025 / Accepted 15 December 2025}

  \abstract{
 The void-galaxy cross-correlation function is a powerful tool to extract cosmological information. Through the void-galaxy cross-correlation function, cosmic voids -- the underdense regions in the galaxy distribution -- are used for refined deductions of the Universe’s content by correcting apparent geometric distortions. This study proposes a novel procedure for optimally extracting the Alcock-Paczy\`nski (AP) signal from cosmic voids through a cosmological reconstruction technique. Employing cosmological reconstruction, specifically using the Zel’dovich approximation, we estimate the true positions of galaxies from their redshift-space locations, reducing distortions introduced by peculiar velocities. Unlike previous analyses, we identify voids and measure the void-galaxy cross-correlation function directly in reconstructed space. 
  This approach enables us, for the first time, to include in our analysis small nonlinear voids, typically discarded in previous studies, thus enhancing the statistical power of void studies and significantly improving their cosmological constraining power. Reconstruction is particularly effective even at small scales for voids, due to their clean and dynamically simple environment. This ability to recover information encoded on small scales significantly enhances the precision of the analysis, leading to a $\sim 23\%$ improvement in the constraints on the AP parameters compared to previous methods where the analysis is performed in redshift space and, consequently, to a better estimate of the derived cosmological parameters. Our analysis also includes a comprehensive set of consistency checks, demonstrating its robustness. We expect this methodology to yield a substantial gain in constraining power when applied to data from modern large-scale structure surveys.}


   \maketitle

\section{Introduction}
Cosmic voids, vast underdense regions that populate the large-scale structure of the Universe, offer a unique and insightful laboratory for cosmology \citep{pisani_2019, kreisch_2019, moresco_2022}. Due to the statistically isotropic nature of structure formation, these initially irregular underdensities evolve toward an approximately spherical shape as matter evacuates their interiors under gravitational dynamics. Although individual voids deviate from sphericity, the process of rescaling by the effective radii and stacking multiple voids yields an average spherically symmetric profile that robustly characterizes their global properties. This averaged spherical structure provides a powerful tool for cosmological analysis. In particular, it enables the application of the Alcock-Paczy\`nski (AP) test \citep{alcock_1979}, which relies on the fact that the observed void profile is obtained by converting galaxy positions observed through directly measurable quantities, i.e., angular coordinates ($\mathbf{\Theta} = (\theta,\phi)$), and redshift ($z$), into comoving distances under an assumed cosmological model. If the fiducial model differs from the true one, the inferred galaxy distribution is systematically distorted, breaking the isotropy of stacked voids. By measuring such distortions in the stacked void profile, the AP test allows us to extract valuable information about the expansion history and geometry of the Universe, as well as the dark energy equation of state. \\
The AP test has been successfully considered for various cosmic structures expected to exhibit statistical isotropy. Examples include the Lyman-$\alpha$ absorption features in the spectra of neighboring quasars \citep{hui_1999, eriksen_2005}, measurements of the auto-correlation function of brightness temperature in 21-cm maps of the epoch of reionization \citep{nusser_2005}, inspections of the Baryon acoustic oscillation (BAO) peak in the anisotropic galaxy 2-point correlation function
 \citep[see e.g.][]{percival_2010}, and full-shape analyses of the same statistics \citep{marulli_2012}.

The idea of applying the AP test to cosmic voids was first proposed by \citet{ryden_1995}, although its implementation on individual voids proved challenging due to their complex and irregular shapes. This limitation was addressed by \citet{lavaux_2012}, who introduced the concept of stacking voids to average out shape irregularities and enhance the statistical power of the test. The first practical applications of the AP test to stacked voids were carried out by \citet{sutter_2012} and \citet{mao_2017}. These developments have led to a shift in methodology, favoring the analysis of AP distortions with the void-galaxy cross-correlation function (VGCF) -- a statistical measurement of how galaxies are distributed around void centers -- which effectively captures the anisotropic features of stacked voids and offers an accurate representation of their average shape. The AP test represents a powerful and well-established method for probing cosmological parameters through the geometric distortions. However, a significant challenge arises from the fact that the AP distortions are not the only ones involved. Indeed, when redshifts are used as proxies for distance, Redshift space distortions (RSD), caused by having neglected the contribution of galaxy peculiar velocities to the observed redshift when converting $z$ into distances, also affect the observed structure of voids. These dynamical distortions are degenerate with the AP ones, making it necessary to disentangle the two effects. The impact of these dynamical distortions is typically addressed by modeling together the effect of RSD and the AP on the measured VGCF \citep[see e.g.][]{nadathur_2020,hamaus_2020, woodfinden_2022, radinovic_2023, hamaus_2022, verza_2024roman}. However, since the two effects are partly degenerate, we need to model RSD accurately to perform an AP test. To address this problem, several RSD models of increasing complexity have been proposed \citep[see e.g.][]{paz_2013,hamaus_2014b,cai_2016,nadathur_2018_accuratemodel, hamaus_2022}. Although these models were sufficient for cosmological analyses of past datasets, such as BOSS \citep{boss_dr12}, the increased statistics coming with data from the next-generation stage IV spectroscopic redshift surveys such as DESI \citep{DESI_Collaboration_2022}, Euclid \citep{mellier_2024_euclid}, and Roman \citep{spergel_2015,dore_2019} requires continual improvements in theoretical modeling to match the unprecedented quality of the data. Two approaches are possible. The first is to account for nonlinear effects in the VGCF by either introducing additional nuisance parameters to empirically model them, as in \citet{hamaus_2020}, or by incorporating physically motivated nonlinear models directly into the likelihood analysis, as in \citet{nadathur_2018_accuratemodel}.
The alternative approach is to apply a theoretically justified transformation to the data to minimize the impact of nonlinear effects. In this work, we adopt the latter, commonly known as cosmological reconstruction.
More specifically, we use the Zel'dovich approximation \citep{zeldovich_1970} to estimate the galaxies' displacement field from their apparent positions in redshift space using first order Lagrangian perturbation theory. This allows us to remove the peculiar velocity component and restore their real-space positions, effectively eliminating RSDs.

While more sophisticated techniques could be used to account for nonlinear effects \citep[see e.g.][]{nusser_2000, sarpa_2021, veena_2022, courtois_2023}, we chose here to rely on the Zel'dovich method, as it is generally considered the standard and has proven to be a robust and reliable methodology in galaxy clustering analyses \citep{padmanabhan_2012, gilmarin_2020, DESI_BAO_2024}.

The cosmological reconstruction approach offers two main advantages. First, by removing RSD and their anisotropies, it allows voids to be identified in real space rather than redshift space, significantly reducing the likelihood of selection effects due to being in redshift space \citep{correa_2022}. Second, smaller cosmic voids are particularly susceptible to nonlinear dynamics that cannot be fully captured by linear RSD models. As a result, most RSD-based analyses \citep[see e.g.][]{hamaus_2020} adopt conservative selection criteria to exclude these smaller voids, which, however, reduces the statistical power of the sample, in particular since small voids are the most abundant. Since the Zel'dovich reconstruction accounts for mildly nonlinear effects, it enables the analysis of smaller-scale voids that are typically excluded in linear RSD analyses, effectively increasing the void sample and reducing statistical errors. 
In line with this, previous studies such as \citet{nadathur2019zeldovich}, \citet{woodfinden_2022}, and \citet{radinovic_2023} have also applied reconstruction techniques, specifically based on the Zel'dovich approximation, to galaxy catalogs prior to void identification, with the aim of mitigating the nonlinearities and selection effects arising in the void-finding process.
However, in those studies, the subsequent analysis was performed by correlating void positions in (reconstructed) real space with galaxy positions in redshift space, eventually using RSD models to compare these VGCF measurements with theoretical predictions. This approach overcomes the challenge of identifying voids in redshift space but does not resolve the limitations associated with modeling RSDs in the void profile, with the consequence of not fully disentangling the two effects. Additionally, this hybrid approach (voids in reconstructed space and galaxies in redshift space) makes the modeling less straightforward. In this work, we aim at conducting the entire analysis in (reconstructed) real space, cross-correlating void and galaxy positions in reconstructed space. Additionally, we aim to investigate whether this approach 
can be successfully applied to small-scale voids, thereby increasing both the statistical power of this  strategy and the precision of the final constraints, in comparison to the standard analysis performed in redshift space \citep[see e.g.][]{hamaus_2020, hamaus_2022, verza_2024roman}. 
Ultimately, in this work we focus on the Zel’dovich reconstruction approach, while a broader comparison with other reconstruction-based methodologies previously mentioned lies beyond the scope of this study and is left for future exploration.

The layout of this article is as follows.
In Section~\ref{sec: theoretical background} we briefly review the AP and RSD effects, describe the VGCF model adopted in this work, and briefly review the Zel'dovich reconstruction method. In Section~\ref{sec: data} we describe the simulated galaxy catalogs that we used for our analyses, while in Section~\ref{sec: likelihood} we describe the likelihood that we use to perform the analysis, together with the estimator of the VGCF and the respective covariance matrix. The results are presented in Section~\ref{sec: results} and discussed in the conclusions (Section~\ref{sec: conclusions}).

\section{Theoretical background}
\label{sec: theoretical background}

In cosmology and specifically in clustering analyses, we typically use the observed redshift, $z_\mathrm{obs}$, as a proxy for a galaxy's distance. This, combined with the observed angular coordinates, $\mathbf{\Theta} = (\theta,\phi)$, allows us to use galaxy redshift surveys to map the distribution of matter in the Universe. 

The redshift $z_\mathrm{obs}$ is converted into comoving distances via: 
\begin{equation}
\label{eq: d(z)}
    d(z) = \int_0^{z_\mathrm{obs}} \frac{c}{H(z')} \, dz',
\end{equation}
where $c$ is the speed of light, and $H(z)$ is the Hubble function, which describes the expansion history of the Universe and depends on the assumed cosmological model through the density parameters $\pmb{\Omega}_i$ of all the components that contribute to the total energy-mass budget of the Universe, and on the dark energy equation of state. 
The distance inferred from the redshift may systematically differ from the true distance for two reasons. First, the assumed (or fiducial) cosmological model may differ from the true one. Adopting an incorrect Hubble function $H(z)$ will systematically bias the distance estimate $d(z)$. This results in the introduction of the characteristic Alcock-Paczi\'nski effect, which distorts the 3D mapping of the galaxy distribution. Second, galaxies' peculiar velocities also contribute to the observed redshift $z_\mathrm{obs}$ through the Doppler effect.
While smaller in magnitude than the cosmological redshift, neglecting this contribution introduces a systematic bias in the estimated distance, as for the AP test.
Moreover, only the radial component of the peculiar velocity affects the Doppler shift, the inferred mass density field therefore lacks the expected statistical isotropy. Both effects introduce distortions in the 3D mapping from redshift to distance. However, since these distortions are directly linked to cosmological parameters, their detection enables comparison with theoretical models to constrain the matter content $\pmb{\Omega}_i$ and the growth rate of structures in the Universe $f(z)$.

\subsection{Alcock-Paczy\'nski distortions}
\label{subsec: ap distortions}

In our analysis, we measure the separation between a galaxy at $(d(z_1), \Theta_1)$ and a cosmic void center at $(d(z_2), \Theta_2)$, both identified in the same spectroscopic sample.
Assuming the distant observer approximation and using the void center as the line-of-sight (LOS) \citep{hamaus_2020}, we decompose the void-galaxy separation vector $\mathbf{r}$ into parallel and perpendicular components, such that $r = \sqrt{r_\parallel^2 + r_\perp^2}$. Using Eq. \eqref{eq: d(z)}, and assuming that $H(z_{1})=H(z_{2})$ and $D_\mathrm{A}(z_{1})=D_\mathrm{A}(z_{2})$, it follows that
\begin{equation}
\label{rpar_rper}
    r_\parallel=\dfrac{c}{H(z)}\delta z \quad r_\perp=D_\mathrm{A}(z)\delta \theta \, ,
\end{equation}
where $\delta z=z_1-z_2$ and $\delta \theta$ denote the redshift and angular separations of the void-galaxy pair, respectively, and $z$ is the mean redshift of the galaxy sample. This assumes that the size of the sample $l$ is much smaller than its distance to the observer $d(z)\ll l$.
$D_\mathrm{A}(z) = d(z)/(1+z)$ is the angular diameter distance and the equality holds in a flat Universe. Since $D_\mathrm{A}(z)$ depends on $d(z)$, and since $d(z)$ itself is a function of $H(z)$, it follows that $D_\mathrm{A}(z)$ also depends on $H(z)$ and thus on the fiducial cosmology. A mismatch between the fiducial and the true cosmology 
systematically biases pairs separations as follows
\begin{equation}
    \label{eq: r_par r_per AP}
    r'_\parallel=r_\parallel\dfrac{H(z)}{H'(z)}\equiv q_\parallel^{-1}r_\parallel \quad r'_\perp=r_\perp\dfrac{D'_\mathrm{A}(z)}{D_\mathrm{A}(z)}\equiv q_\perp^{-1}r_\perp,
\end{equation}
where the primed quantities are evaluated in the fiducial cosmology. 
In the case of voids, that are standard spheres \citep{lavaux_2012}, meaning that they are spherically symmetric but lacking a characteristic scale as the BAO, the AP test constrains only the product $D_\mathrm{A}(z) H(z)$ via the ellipticity parameter $\varepsilon$:
\begin{equation}
    \label{eq: epsilon}
    \varepsilon\equiv\frac{q_\perp}{q_\parallel}=\frac{D_\mathrm{A}(z)H(z)}{D'_\mathrm{A}(z)H'(z)}.
\end{equation}
Deviations from $\varepsilon = 1$ indicate a mismatch between the fiducial and true cosmologies.\\
Our aim is to constrain the AP distortion parameter $\varepsilon$ using the multipoles of the VGCF. To enable stacking across voids of varying size, thus focusing on the average profile, we normalize each void-galaxy separation by the void’s effective radius $R$, defined as the cube root of its volume (see Section~\ref{subsec: void catalogs}). We also account for the AP scaling of $R$ \citep{hamaus_2020} 
\begin{equation}
R = q_\parallel^{1/3} q_\perp^{2/3} R'.
\end{equation}

Additionally, we express the separation vector ${\bf r}$ in terms of its absolute value $r$, and the cosine of its angle with respect to the LOS $\mu = r_\parallel/r$. In these new coordinates, the effect of AP distortions is parametrized as \citep{hamaus_2020} 
\begin{equation}
    \label{eq: s/R AP}
    \frac{r}{R}\left(\varepsilon\right)=\dfrac{r'}{R'}\mu'\varepsilon^{2/3}\sqrt{1+\varepsilon^2(\mu'^{-2}-1)},
\end{equation}
\begin{equation}
    \label{eq: mu AP}
    \mu(\varepsilon)=\dfrac{\text{sgn}(\mu')}{\sqrt{1+\varepsilon^2(\mu'^{-2}-1})} \, ,
\end{equation}
where primed quantities are estimated using the fiducial cosmological model.\\
For simplicity of notation, we will henceforth denote by $r$ the void–galaxy separation normalized by the effective void radius, i.e., $r \equiv r / R$.

\subsection{Redshift space distortions}
\label{subsec: rsd}

The observed redshift $z_\mathrm{obs}$ results from the combination of the cosmological redshift, $z_\mathrm{c}$ of a galaxy generated by the Hubble expansion, and a Doppler effect due to the LOS component of its peculiar velocity, $z_\mathrm{d}=v_\parallel/c$. The observed redshift can be expressed as 
\begin{equation}
    \label{eq: 1+z}
    1+z_\mathrm{obs}=(1+z_\mathrm{c})(1+z_\mathrm{d}).
\end{equation}
If $z_\mathrm{obs}$ is used as a distance proxy, meaning it is employed in Eq. \eqref{eq: d(z)}  to estimate the galaxy distance, a systematic error is introduced by a non-vanishing 
 $z_\mathrm{d}$. This effectively displaces the estimated position of the galaxy relative to its true position along the LOS. The resulting redshift space distortions break the statistical isotropy of the galaxy sample.
A simple relation between the true void-galaxy separation vector, $\mathbf{r}$, and the one estimated from the observed redshift, $\mathbf{s}$, can be obtained in the limit $z_\mathrm{d}\ll z_\mathrm{c}$, by assuming that the cosmological redshift of the void center is the same as that of the galaxy in the pair:
\begin{equation}
    \label{eq: mapping_s(r) only s->r}
    \mathbf{s}=\mathbf{r}+\frac{1+z}{H(z)}\mathbf{u}_\parallel \, ,
\end{equation}
where $\mathbf{u}_\parallel$ is the LOS component of the relative void-galaxy peculiar velocity.

A relation between the peculiar velocity vector and that the separation vector to the center of the void can be obtained from the continuity equation, imposing local mass conservation, by assuming linear Eulerian perturbation theory \citep{peebles_1980}. Its expression for a spherically symmetric void is \citep{peebles_1980, cai_2016}:
\begin{equation}
    \label{eq: u(r)}
    \mathbf{u}(\mathbf{r})=-\dfrac{f}{3}\dfrac{H(z)}{1+z}\Delta(r)\mathbf{r},
\end{equation}
where $f$ is the linear growth rate of the mass density fluctuations, and $\Delta(r)$ the density contrast $\delta$ averaged within a comoving radius $r$ 
\begin{equation}
    \label{eq: Delta(r)}
    \Delta(r)=\frac{3}{r^3}\int_0^r\delta(r')r'^{2}dr' \, .
\end{equation}
In order to evaluate the redshift-to-real space mapping we plug Eqs.~\eqref{eq: u(r)} and \eqref{eq: Delta(r)} into Eq.~\eqref{eq: mapping_s(r) only s->r}, obtaining 
\begin{equation}
\label{eq: mapping_s(r)}
\begin{aligned}
    r_\parallel &= \dfrac{s_\parallel}{1 - \dfrac{f(z)}{3} \Delta(r)} \\
    r_\perp &= s_\perp
\end{aligned}
\end{equation}

\subsection{Modeling the void-galaxy cross-correlation function.}
\label{subsec: modeling ccf}

The void-galaxy cross-correlation function (VGCF), $\xi(r)$, quantifies the excess probability of finding a galaxy at a comoving distance $r$ from the center of a void, compared to a random distribution of objects, with no clustering properties. In a statistically isotropic universe, $\xi$ would depend only on the modulus of the separation vector. In the presence of RSD or AP distortions, however, this is no longer the case: their effects on the VGCF must be properly modeled.
Since RSD (dynamical) and AP (geometrical) distortions are independent, we model them separately. We first describe RSD in the true cosmology and later introduce AP effects. From now on, the pair separation vectors in redshift and real space, $r$ and $s$, are understood as dimensionless quantities obtained by dividing the physical separation by the voids' radii.\\
The RSD effect can be modeled using linear perturbation theory to express the Jacobian of the redshift-to-real space coordinate transformation, Eq.~\eqref{eq: mapping_s(r)}. The resulting expression, at linear order, for the VGCF is \citep{cai_2016}:
\begin{equation}
    \label{eq: xi model RSD}
    \xi(s,\mu)= \xi(r) + \frac{\beta}{3}\bar{\xi}(r)+\beta\mu^2\left[\xi(r)-\bar{\xi}(r) \right] \,
\end{equation}
where we assume a linear bias $b$ relation that links the observed galaxy density field to the underlying matter density field, obtaining a relation between the VGCF and the void density profile of the form $\xi(r) = b\delta(r)$, and consequently $\bar{\xi}(r) = b\Delta(r)$, and the distortion parameter is $\beta\equiv f/b$. The mapping, Eq.~\eqref{eq: mapping_s(r)}, simplifies to:
\begin{equation}
\label{eq: mapping_s(r) beta}
\begin{aligned}
    r_\parallel &= \dfrac{s_\parallel}{1 - \dfrac{\beta}{3} \bar{\xi}(r)} \\
    r_\perp &= s_\perp
\end{aligned}
\end{equation}

The last three equations fully describe the RSD in the linear regime. The AP distortions can now be introduced by applying the coordinate transformations in Eqs.~\eqref{eq: s/R AP} and \eqref{eq: mu AP}:
\begin{equation}
    \label{eq: xi model AP RSD 1}
    \xi(s,\mu)= \xi(r) + \frac{\beta}{3}\bar{\xi}(r)+\beta\mu^2\left[\xi(r)-\bar{\xi}(r) \right] \Bigg|_{r=r(\varepsilon), \ s=s(\varepsilon),\  \mu=\mu(\varepsilon)}\, ,
\end{equation}

To highlight the anisotropy of the VGCF, the function $ \xi\left(s, \mu\right)$ can be decomposed into multipoles by use of the Legendre polynomials $P_\ell$ of order $\ell$, using
\begin{equation}
    \label{eq: xi_elle}
    \xi_\ell(s)=\frac{2\ell+1}{2}\int_{-1}^{1} \xi\left(s, \mu\right) P_\ell(\mu)\mathrm{d}\mu\ .
\end{equation}

In this analysis, we use the first three even multipoles of the VGCF: the monopole $\xi_0(s)$, the quadrupole $\xi_2(s)$, and the hexadecapole $\xi_4(s)$. While odd multipoles are expected to vanish due to statistical isotropy, only the monopole and quadrupole are generally predicted to be non-zero in redshift space.  
Their interpretation in terms of void density profile is as follows. The monopole corresponds to the average number density profile of tracers within a distance $s$ from the void center. The quadrupole is expected to be zero for a statistically isotropic sample and either negative or positive in the presence of RSD and/or AP distortions. A positive quadrupole is present when the density profile is elongated along the LOS. A negative quadrupole indicates that the density profile of the mean void is squeezed along the LOS direction. The inclusion of the hexadecapole, although not strongly motivated by theory, is intended to capture potential additional distortions or systematics not accounted for by the model. A non-zero hexadecapole indicates a violation of statistical isotropy. However, its interpretation in terms of void profile is less straightforward.

\subsection{Beyond the linear VGCF model}
\label{subsec: issues with the modeling}

The range of validity of the linear VGCF model has been widely discussed in recent literature. While \citet{schuster_2022} argue that the linearized continuity equation Eq.~\eqref{eq: u(r)} remains accurate down to relatively small scales, \citet{nadathur_2018_accuratemodel} highlight its limited applicability in redshift space. These limitations have motivated several improvements to the linear model. For example, \citet{hamaus_2020} introduced nuisance parameters to marginalize over systematic effects, such as nonlinear corrections, spurious voids, and selection biases (see below). Conversely, \citet{nadathur_2018_accuratemodel} incorporated explicit correction terms to account for deviations from linear theory.

However, inaccuracies in the modeling are not the only factors affecting void analyses. A significant contribution to the total error budget arises from the void identification process itself, discussed in Section~\ref{subsec: void catalogs}. Most commonly used void finders are based on topological criteria applied to the spatial distribution of discrete tracers, such as galaxies, rather than on the underlying dynamical field. Although these methods are straightforward to apply in both observational and simulated data, they suffer from two key limitations.

The first is the selection of spurious voids, Poisson fluctuations mistaken for genuine underdensities, which reduces the purity of the void catalog and suppresses the amplitude of the VGCF quadrupole moment (see \citealp{cousinou_2019}; \citealp{hawken_2020}; \citealp{aubert_2022}). The second concerns the violation of pair count conservation between real and redshift space, caused by the fragmentation of voids in redshift space, which introduces additional systematics \citep{correa_2022}. To mitigate this effect, \citet{nadathur2019zeldovich} proposed the use of a reconstruction algorithm to partially remove RSD and identify voids in the reconstructed real-space density field, thereby recovering a more consistent void population across the two spaces, and correlated such voids with galaxies in redshift space.\\

Another potential source of disturbance in the model is the assumption of linear galaxy bias, which may not hold in the inner regions of cosmic voids. This issue is thoroughly discussed by \cite{nadathur_2018_accuratemodel, hamaus_2020} in relation to the VGCF analysis, and additionally by \cite{pollina_2017, pollina_2019, contarini_2019, verza_2022}. 
In this work, we adopt a phenomenological approach and use the VGCF model based on an empirical modification of Eqs. \eqref{eq: xi model RSD} and \eqref{eq: mapping_s(r) beta}, proposed by \cite{hamaus_2022}.
This model introduces two additional nuisance parameters, $\mathcal{M}$ and $\mathcal{Q}$, compared to the linear RSD model for voids proposed by \citet{cai_2016}, and modifies the RSD terms by altering the coefficients that govern the amplitude of the distortions (see \citealp{hamaus_2022} for details):

\begin{equation}
    \label{eq: xi model RSD MQ}
    \xi^s(s,\mu)=\mathcal{M} \biggl\{  \xi(r) + \beta\bar{\xi}(r)+2\mathcal{Q}\beta\mu^2\left[\xi(r)-\bar{\xi}(r) \right] \biggr\} \ .
\end{equation}

The nuisance parameters $\mathcal{M}$ and $\mathcal{Q}$ are introduced by \citet{hamaus_2020} to account for potential systematics, including potential deviations from the linear theory. This would include deviations in the dynamics and the bias parameter, as well as the potential presence of spurious voids in the sample, which may reduce the amplitude of both the monopole and quadrupole moments of the VGCF \citep{cousinou_2019}.
The monopole-like parameter $\mathcal{M}$ regulates the overall amplitude of the VGCF and is intended to account for all factors that may influence the magnitude of the real-space void-galaxy clustering.
The parameter $\mathcal{Q}$ multiplies the $\mu^2$ term, thereby regulating the quadrupole moment. Its role consists of mitigating the loss of amplitude due to spurious voids.\\

Finally, the model requires the real-space VGCF, $\xi(r)$, along with its volume-averaged counterpart, $\bar{\xi}(r)$, which we estimate phenomenologically from the mean real-space VGCF monopole measured across a suite of 100 mock catalogs (described in Section~\ref{subsec: mock}). While this choice provides a robust baseline for our analysis, a more data-driven treatment based on alternative prescriptions for modeling $\bar{\xi}(r)$ is left to future work \citep[see e.g.][]{pisani_2014, hawken_2017}. Furthermore, Eq. \eqref{eq: xi model RSD MQ} already requires the knowledge of the argument $r$ of $\bar{\xi}(r)$. When performing the analysis in redshift space (not in reconstructed space) we need to account for this, therefore we use the method proposed by \citet{hamaus_2020} to evaluate this quantity via iteration using Eq.~\eqref{eq: mapping_s(r) beta}: we start with using $\bar{\xi}(s)$ as initial guess for $\bar{\xi}(r)$, and iteratively calculate $r_\parallel$ and $\bar{\xi}(r)$, until convergence is reached. As with any phenomenological model, it includes uncertainties that primarily affect the amplitude of the VGCF. For this reason, such uncertainties are expected to be absorbed by the parameter $\mathcal{M}$. We incorporate the AP effect by applying the coordinate transformation described in Eqs.~\eqref{eq: s/R AP} and \eqref{eq: mu AP}, effectively introducing an explicit dependence of the coordinate $r$, $s$, $\mu$, on the AP distortion parameter $\varepsilon$ in Eq.~\eqref{eq: xi model RSD MQ}.

\subsection{Cosmological reconstruction}
\label{subsec: reconstruction}

To perform the reconstruction, we apply the Zel'dovich approximation \citep{zeldovich_1970} to describe the dynamics of a self-gravitating system of mass tracers and use the publicly available \textsc{LinearVelocity} code \footnote{\url{https://gitlab.com/esarpa1/linear_velocity}} to recover galaxies' real-space positions from their observed distribution in redshift space. The code implements the \verb|MultiGridReconstruction| method \citep{white_notes} as integrated in the Python package \textsc{pyrecon}\footnote{\url{https://github.com/cosmodesi/pyrecon/tree/main}}. The inputs to the reconstruction are the angular positions and redshifts of a set of galaxies in a spectroscopic survey, which trace the matter density field. Additionally, the inputs include the selection function in the form of a random catalog of unclustered objects, a fiducial value for the linear galaxy bias $b$, a fiducial value for the linear growth rate of cosmic structures $f$, and the radius of the Gaussian filter $\mathcal{S}(k)$, which is applied in Fourier space to suppress nonlinear features in the input matter density field that cannot be captured by the Zel'dovich approximation. Moreover, the user can specify the number of grid cells $N_\mathrm{cell}$ over which the peculiar velocity field will be evaluated.

The reconstruction is performed in several steps. In the first step, the positions of galaxies are interpolated onto a cubic grid to estimate their number density, $\rho_\mathrm{obs}$. The cube is large enough to encompass the survey volume. The procedure is repeated to estimate, at the same positions, the number density of the random objects $\rho_\mathrm{r}$. 
The overdensity field is then computed as:
$\delta_\mathrm{obs} = 
\frac{\rho_\mathrm{obs}}{\rho_\mathrm{r}} - 1$.
Since a linear bias is assumed, the mass overdensity can be estimated by simply rescaling the galaxy overdensity by the bias factor $b$.  In cells where $\rho_\mathrm{r} = 0$, the overdensity is set to zero to avoid numerical instabilities. In the second step, a Gaussian filter is applied to the overdensity field. Since this operation corresponds to a simple multiplication in Fourier space, we first Fourier transform the overdensity field and multiply it by the Gaussian filter $\mathcal{S}(\mathbf{k}) = e^{-k^2R_\mathrm{s}^2/2}$,  where $R_\mathrm{s}$ is the smoothing scale. Finally, we inverse Fourier transform the smoothed field back to configuration space. In the third, and central, step it evaluates the displacement field $\boldsymbol{\psi}_\mathrm{RSD}$ by solving the system of equations \citep{nusser_1994}:
\begin{eqnarray}
\label{eq:ZA_disp_sol}
\boldsymbol{\nabla}\cdot\boldsymbol{\psi}_\mathrm{tot} + \frac{f}{b}\boldsymbol{\nabla}\cdot\left[\frac{\mathbf{s}}{s^2}\left(\mathbf{s}\cdot \boldsymbol{\psi}_\mathrm{tot}\right)\right] &=& \frac{\delta_\mathrm{obs}}{b}, \\
\boldsymbol{\psi}_\mathrm{RSD} &=& \frac{f}{b}\boldsymbol{\nabla}\cdot\left[\frac{\mathbf{s}}{s^2}\left(\mathbf{s}\cdot \boldsymbol{\psi}_\mathrm{tot}\right)\right].
\end{eqnarray}

In the first equation, the estimated overdensity field $\delta_\mathrm{obs}$ is used to derive the Zel'dovich displacement field, $\boldsymbol{\psi}_\mathrm{tot}$, which describes the straight orbits back to the initial positions. This displacement field is then used in the second equation to derive a second displacement field, $\boldsymbol{\psi}_\mathrm{RSD}$, evaluated at the observed positions of the galaxies. 
 This system of equations is solved in configuration space using a multigrid V-cycle, where the solution is iteratively transferred between fine and coarse grids to accelerate convergence by correcting low-frequency errors, while damped Jacobi iterations update each grid point from its neighbors to stabilize the relaxation. In the fourth and final step, the displacement field $\boldsymbol{\psi}_\mathrm{RSD}$ is used to shift galaxies from their redshift-space positions to their real-space positions, completing the reconstruction process.

Zel'dovich-based reconstruction techniques, similar to the one presented here, have been extensively applied to the analysis of large-scale structures, particularly to enhance the signal-to-noise ratio of the BAO peak in the galaxy-galaxy correlation function \citep[see e.g.][]{eisenstein_2005, padmanabhan_2012, ross_2017}. 
More recently, these methods have also been employed in void studies \citep{nadathur2019zeldovich, woodfinden_2022, radinovic_2023} (albeit focusing on finding voids in reconstructed space but correlating with galaxies in redshift space, therefore using a hybrid set-up, different to what is considered in this paper). Building upon these works, our study shifts the focus exclusively to the estimation of AP parameters, i.e., to probing the background geometry alone, rather than jointly constraining it with the growth rate of structure, $f$, as in previous analyses. By performing the entire analysis, including void identification, directly in the reconstructed space, we eliminate the need to model RSDs at any stage. This approach effectively disentangles geometric distortions due to the AP effect from those induced by RSDs.

\section{Data}
\label{sec: data}

\subsection{The simulated halo catalogs}
\label{subsec: mock}

To validate our void-galaxy cross-correlation strategy, we used simulated halo catalogs extracted from $N$-body simulations. Given that most recent void analyses have focused on BOSS galaxy data \citep{hamaus_2020, nadathur_2020, woodfinden_2022}, we chose to work with mock catalogs that match the mean redshift, e.g., $\bar{z}\simeq0.51$, and number density, $n(\bar{z})=4\times 10^{-4}\ h^3\text{Mpc}^{-3}$, of the sample used for the void analyses of BOSS data.
Our parent sample consists of 100 independent realizations from the publicly available Quijote $N$-body simulation suite\footnote{\url{https://quijote-simulations.readthedocs.io/en/latest/}} \citep{quijote}, all run with the same flat $\Lambda$CDM cosmology: $h = 0.6711$, $\Omega_\mathrm{m} = 0.3175$, $\Omega_\mathrm{b} = 0.049$, $n_\mathrm{s} = 0.9624$, and $\sigma_8 = 0.834$. Each simulation spans a box of $L_\mathrm{box} = 1000\ h^{-1} \mathrm{Mpc}$ and we consider snapshots at $z_\mathrm{snap} = 0.5$, close to the typical redshift of the BOSS dataset \citep{reid_2016}. Halos are identified using the friends-of-friends (FoF) algorithm. This approach considers only the parent halos -- and not the subhalos -- which correspond to trace the central galaxies while ignoring satellites. This approximation is adequate for describing the luminous red galaxy (LRG) sample extracted from the BOSS data, as these galaxies are expected to reside at the centers of massive halos. To match the number density of the LOWZ and CMASS samples in SDSS-III DR12 \citep{reid_2016}, we applied a mass cut of $m_\mathrm{min} = 10^{13} \ h^{-1} M_\odot$, yielding about $4 \times 10^5$ halos per catalog.

We also computed the fiducial growth rate at $z_\mathrm{snap}$ as $f = \Omega_\mathrm{m}^{0.55} = 0.763$, and determined the effective halo bias to be $b = 1.87 \pm 0.03$, from the measured 2-point correlation function, as detailed in  Appendix~\ref{appendix: bias}. These values yield a distortion parameter $\beta = f/b=0.41$, which is used in the reconstruction process (Section~\ref{subsec: reconstruction}). \\
To build the three types of catalogs used in our analysis, we took the positions of the halo in real space and move them along the LOS to their redshift-space position using Eq. 
\ref{eq: mapping_s(r) only s->r} combined with the peculiar velocities provided by the simulation. We then applied the reconstruction procedure to the redshift-space halo catalog (see Section~\ref{subsec: redshift space catalogs}). As a result, we obtained three distinct types of halo catalogs: one in real space, one in redshift space, and one in reconstructed space. We then apply the void finder to each of these catalogs (see Section~\ref{subsec: void catalogs}), producing the three corresponding void catalogs used in our cross-correlation analyses.

\subsection{The redshift space halo catalogs}
\label{subsec: redshift space catalogs}

To generate the redshift-space catalogs from the real-space data described in Section \ref{subsec: mock}, we adopted the distant observer approximation, aligning the LOS with the Cartesian $z$-axis. The center of the box is placed at a comoving distance $d(z_\mathrm{snap})$, as given by Eq.~\eqref{eq: d(z)}, and distances to all halos are computed accordingly. The observed redshift of each halo is then calculated using Eq.~\eqref{eq: 1+z}, combining the cosmological redshift with the Doppler shift from the LOS component of peculiar velocities, $z_d = \mathbf{v}_\parallel/c$, provided by the simulation.
We used Eq.~\ref{eq: d(z)} to estimate the new comoving distances and their Cartesian coordinates $(x, y, z)$.
Since this transformation can shift some halos beyond the cube boundaries, we trimmed the catalog to exclude those objects. The resulting loss is negligible, affecting only $\sim 0.3\%$ of halos. These redshift-space mocks are used both for the redshift space analysis and as input to the reconstruction algorithm to extract the reconstructed-space catalogs.

\subsection{The reconstructed halo catalogs}
\label{subsec: reconstructed catalogs}
Reconstructed space halo catalogs were obtained by applying the Zel'dovich reconstruction algorithm to the redshift-space halo catalogs. The reconstruction procedure (Section~\ref{subsec: reconstruction}) requires defining a cubic grid of cells to interpolate the halo number density. We set $N_\mathrm{cell}=256$, corresponding to cell size of $3.9\ h^{-1}\mathrm{Mpc}$. The density field is then smoothed with a Gaussian filter of radius $R_\mathrm{s}\sim 5\ h^{-1}\mathrm{Mpc}$ to minimize the residual RSD effects. The choice of $R_\mathrm{s}$ is motivated in the Appendix~\ref{appendix: robustness tests}. As for the number of cells, we verified that its choice has no significant impact on the reconstruction as long as the size of the cell is smaller than the smoothing radius \citep{white_notes}. The fact that the smoothing radius is just 1.5 smaller than that of the smallest void in the sample ($\sim 7.5\ h^{-1}\mathrm{Mpc}$) has no significant impact on our results. The role of the smoothing is to minimize the effects of nonlinearities when removing RSDs, which it does effectively, as shown in the following sections. Importantly, it plays no role in the void detection process, which relies on a Voronoi tessellation of the discrete tracer distribution.
As in the previous section, we removed all halos displaced outside the cube resulting in a loss of $\sim 0.4\%$ with respect to the original catalog in real space.

\subsection{The void catalogs}
\label{subsec: void catalogs}

\begin{figure}[ht]
    \centering
    \includegraphics[width=0.9\linewidth]{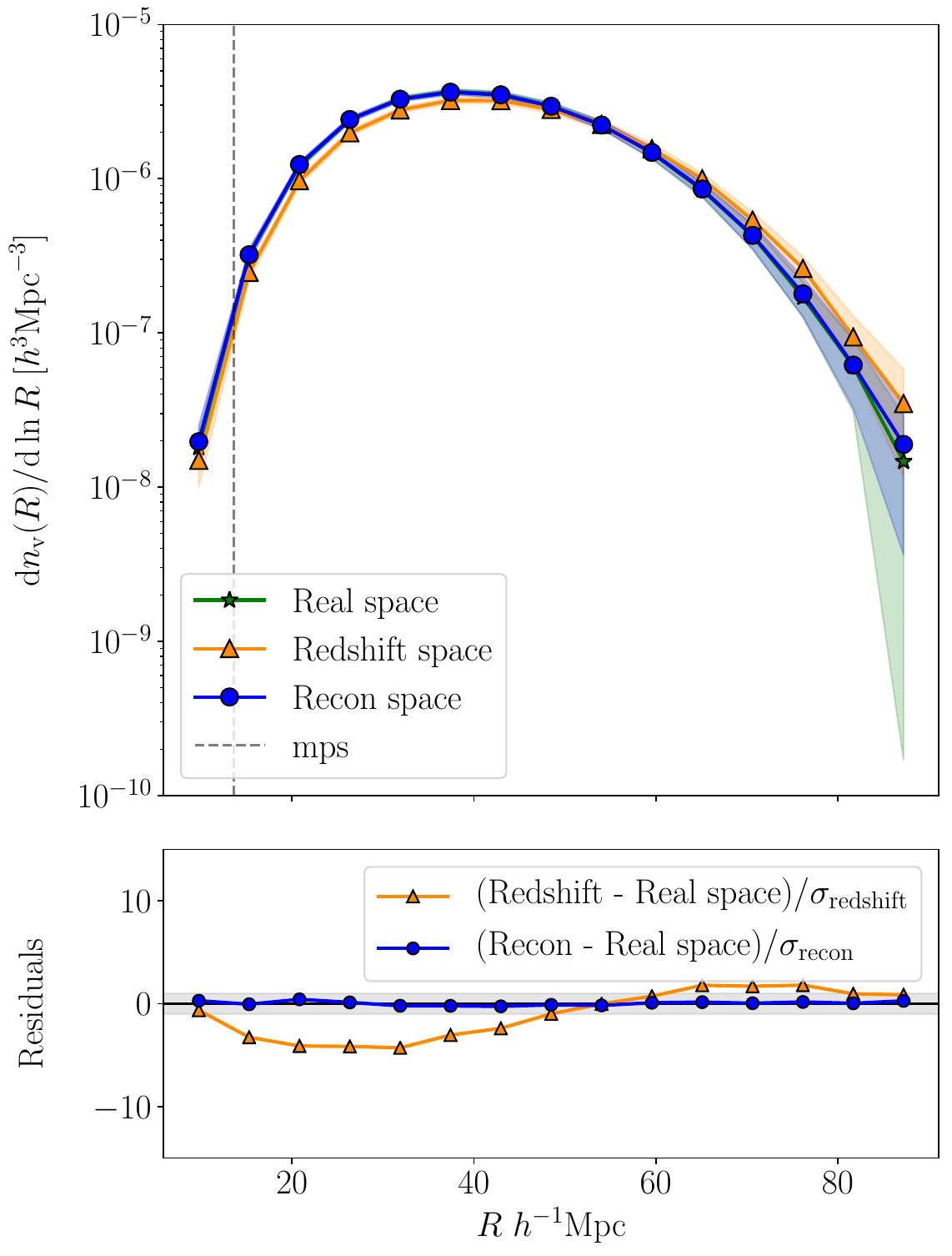}
    \caption{\textit{Top Panel}: number density of voids per unit radius interval, i.e., the void size function, of the {\fontfamily{cmtt}\selectfont VIDE} voids found in real (green stars), redshift (orange triangles), and reconstructed (blue dots) space as a function of their effective
radius $R$, averaged over 100 mocks; shaded bands show the $1\sigma$ standard deviation across mocks. The dashed line marks the mean tracer separation of the halo simulations. \textit{Bottom Panel}: residuals divided by $1\sigma$ standard deviation relative to real space, redshift - real (orange) and reconstructed - real (blue), showing reconstructed space agrees more closely with real space. The grey band indicates the $[-1,1]\ \sigma$ interval. }
    \label{fig: vsf}
\end{figure}

Void analyses identify these structures in a 3-dimensional distribution of mass tracers, in our case simulated dark matter halos. 
Several void identification algorithms have been proposed in the literature, \citep[see e.g. ][for a cross-comparison of various techniques available at that time]{colberg_2008, cautun_2014}.  In this work, we identified voids in the simulated halo catalogs with the publicly available code {\fontfamily{cmtt}\selectfont VIDE} \footnote{\url{https://bitbucket.org/cosmicvoids/vide_public/}} \citep{sutter_2015_vide}, based on the watershed algorithm {\fontfamily{cmtt}\selectfont ZOBOV} \citep{neyrinck_2008}, which has been widely tested and applied in the literature \citep[e.g.][]{hamaus_2020, contarini_2023, verza_2024}. This void finder estimates local densities using Voronoi tessellation and identifies voids as basins in the tracer field. The output is a catalog of voids, each described by its center
\begin{equation}
\label{eq: void centers VIDE}
\textbf{x}_\mathrm{v}=\frac{\sum_j\textbf{x}_jV_j}{\sum_j V_j},
\end{equation}
calculated as the volume-weighted barycenters, and the effective radius
\begin{equation}
\label{eq: Rv VIDE}
R=\left( \frac{3}{4\pi}\sum_j V_j\right)^{1/3},
\end{equation}
defined as the radius of a sphere with equivalent volume. 
The definition of the void center is robust against Poisson density fluctuations, as selected voids are characterized by a large number of tracer particles \citep{hamaus_2020}.\\
We applied \vide to all three types of catalogs (real, redshift, and reconstructed space) across 100 mocks. Fig.~\ref{fig: vsf} on the top panel shows the void size function, i.e., the abundance of voids as a function of their effective radius calculated as the number density of voids $n_\mathrm{v}$ per unit radius interval, in real (green stars), redshift (orange triangles), and reconstructed (blue dots) space. The solid lines represent the mean across 100 mocks, while the shaded regions indicate the standard deviation. The results in real and reconstructed space are in good agreement within errors, whereas redshift space shows a slight mismatch, as can be better appreciated by looking at the difference with the real space void size function, normalized by the $1\sigma$ error, represented in the bottom panel of Fig.~\ref{fig: vsf}. The discrepancies between redshift and real space are particularly prominent at intermediate scales $[20,40]\ h^{-1}$Mpc where the voids are more abundant, while residuals between real and reconstructed space do not show a dependence on the size, always being within the $1\sigma$ interval (grey band).
This suggests that the reconstruction procedure successfully recovers the statistical properties of the void distribution, including both their abundance and size, while redshift space distortions affect their identification. A possible explanation is that voids squished along the LOS in real space are more likely to not be selected in real space while identified in redshift space, as discussed in \citet{correa_2022}. Additionally, void finders identify spurious, shallow voids due to shot noise from the discrete tracer distribution, commonly referred to as Poisson voids \citep{neyrinck_2008, cousinou_2019}. These will impact void statistics such as the void size function and the VGCF. A standard mitigation is to exclude small voids below a fixed multiple of the mean tracer separation (mps) as in \citet{hamaus_2020}, with values typically around 3 times the mps. However, such cuts are somewhat arbitrary and not fully effective at eliminating contaminants, while considerably reducing the sample size \citep{cousinou_2019}. For this reason, we chose to retain all identified voids and test the robustness of our analysis by comparing results with and without small voids (see Section~\ref{subsec: void radii}). A more detailed investigation into the impact of spurious voids is left for future work.

\section{Likelihood analysis}
\label{sec: likelihood}

In this section, we present the likelihood analyses designed to identify the set of model parameters that best represent the observed VGCF, addressing the key scientific question of this study: whether we can improve the precision of cosmological parameter constraints by including small-scale voids from reconstructed catalogs. 

A central goal of this work is to critically evaluate the performance of the reconstruction procedure within the VGCF framework, particularly in mitigating RSD that can obscure the AP signal. Beyond this, we aim to quantify the benefits of conducting VGCF analyses in reconstructed space compared to analyses performed directly in redshift space, thereby assessing how reconstruction improves the robustness and precision of void-based cosmological inferences, especially on small scales.

All these analyses compare the same type of data vector, consisting of the three even VGCF multipoles, but use different parameter vectors. 

\subsection{Estimator}
\label{subsec: estimator}

The multipoles of the VGCF are obtained using the Davis-Peebles (DP) auto-correlation function estimator \citep{davis_peebles}. The DP estimator is based on counting object pairs at a given separation and comparing them with pair counts performed in a population of synthetic objects distributed over the same volume and sharing the same selection effects as the real ones but with no intrinsic spatial correlation properties, the so-called random sample.

The modified DP estimator, for the void-halo cross-correlation function is defined as: 
\begin{equation}
    \label{eq: DP}
    \xi^\mathrm{DP}(r,\mu)=\frac{n_\mathrm{R}}{n_\mathrm{h}}\frac{\mathcal{D}_\text{v}\mathcal{D}_\mathrm{h}(r,\mu)}{\mathcal{D}_\text{v}\mathcal{R}_\mathrm{h}(r,\mu)}-1,
\end{equation}
where $\mathcal{D}_\text{v}\mathcal{D}_\mathrm{h}$ are the void-halo (or more generally void-tracer) pair counts at radial $r$ and angular $\mu$ separation, and $\mathcal{D}_\text{v}\mathcal{R}_\mathrm{h}$ are the void-random pair counts. The quantities $n_\mathrm{R}$ and $n_\mathrm{h}$ represent the total number counts of the random objects and of halos respectively. In Eq.~\eqref{eq: DP} $r$ is a dimensionless quantity representing the physical separation between the center of the void and the halo divided by the void effective radius $R$. 
In this analysis the pair counts are evaluated in 25 equally-spaced bins in the range $[0,3]$ while $\mu$ is evaluated in $100$ bins between $[0,1]$. In our analysis, we computed the cross pair with the random catalog by generating a single catalog of synthetic objects uniformly distributed across the volume, with a number density 20 times higher than that of the halos. This random catalog is used for all cross-correlation analyses, as they share the same geometry and mean halo density. \\
Once the VGCF is evaluated, the multipoles can be computed using Eq.~\eqref{eq: xi_elle}. Particular attention is given to the quadrupole, as it serves as a crucial diagnostic tool for understanding the effects of RSD. In real space, where no RSD are present, the quadrupole signal is expected to vanish due to the spherical symmetry of the stacked void profile. Thus, a zero quadrupole signal is used as a benchmark to assess the accuracy of the reconstruction procedure within the appropriate cosmological model.
In addition to the impact of RSD on the quadrupole shape, AP geometric distortions cause the quadrupole to deviate from zero. The direction of the deviation, compression or elongation, depends on whether the parameter $\varepsilon$, defined by Eq.~\eqref{eq: epsilon}, is lower or greater than 1. As we use the fiducial cosmology of the simulation in the distance estimate, no genuine AP-induced distortion is expected, and $\varepsilon$ should therefore be unity.

\subsection{Likelihood}
\label{subsec: likelihood}

We performed two likelihood analyses based on different data vectors and models, depending on whether we consider data in redshift space (with RSDs) or in real/reconstructed space (without RSDs). In both cases, the data vector is constructed from the measured multipoles of the VGCF in the corresponding space and compared to the theoretical model in Eq.~\eqref{eq: xi model RSD MQ}. In redshift space, the model parameter vector is $\pmb{\Theta} = (\varepsilon, \beta, \mathcal{M}, \mathcal{Q})$, while in real and reconstructed space, $\mathcal{Q}$ is fixed to 1 and $\pmb{\Theta} = (\varepsilon, \beta, \mathcal{M})$. This choice is supported by the fact that $\mathcal{Q}$ is poorly constrained in the absence of RSDs. Fixing $\mathcal{Q}$ while keeping $\beta$ free allows the latter to serve as an indicator of the effectiveness of RSD removal in reconstructed space.

The Gaussian likelihood function $\mathcal{L}\left(\pmb{\xi}|\pmb{\Theta}\right)$ of the data vector $\pmb{\xi}$, given the model parameter vector $\pmb{\Theta}$ (adapted according to the space considered), is defined as:
\begin{equation}
    \label{eq: likelihood}
    \ln \mathcal{L}\left(\pmb{\xi}|\pmb{\Theta}\right) = -\frac{1}{2}\sum_{i,j}\left[\pmb{\xi}(r_i)-\pmb{\xi}(r_i|\pmb{\Theta})  \right]\text{Cov}[\pmb{\xi}]_{ij}^{-1} \left[\pmb{\xi}(r_j)-\pmb{\xi}(r_j|\pmb{\Theta})  \right].
\end{equation}
We assume uniform priors on the model parameters, with $\varepsilon \in [0,2]$ (well beyond the percent precision required by modern surveys), $\beta \in [-1,1]$, $\mathcal{M} \in [-10,10]$, and (in the redshift-space case only) $\mathcal{Q} \in [-10,10]$.

To perform the parameter estimation, we simultaneously fit the 100 mock catalogs described in Section~\ref{sec: data} and adopt the approach described in \cite{veropalumbo_2022}. For each mock, we compute its individual likelihood using the corresponding measured VGCF multipoles and its own jackknife covariance matrix. The total likelihood is then constructed as the product of the individual likelihoods:
\begin{equation}
    \ln \mathcal{L}_\mathrm{tot}(\pmb{\Theta}) = \sum_{n=1}^{N} \ln \mathcal{L}^{(n)}(\pmb{\xi}^{(n)}|\pmb{\Theta}),
\end{equation}
where $N = 100$, and $\mathcal{L}^{(n)}$ denotes the likelihood computed from the $n$-th mock.

Since the total likelihood accumulates information across all mocks, it would naturally lead to small uncertainties, effectively as if the data had been averaged over 100 realizations. 
To ensure that the combined analysis retains the same statistical weight as that of a single mock, our goal in this work, we rescale the total log-likelihood by multiplying it by $N$. 
This guarantees that the resulting constraints are equivalent to those one would obtain from analyzing a single mock: 
\begin{equation}
    \ln \mathcal{L}_\mathrm{eff}(\pmb{\Theta}) = N \cdot \ln \mathcal{L}_\mathrm{tot}(\pmb{\Theta}).
\end{equation}

\begin{figure*}[!ht]
    \centering
    \includegraphics[width=\linewidth]{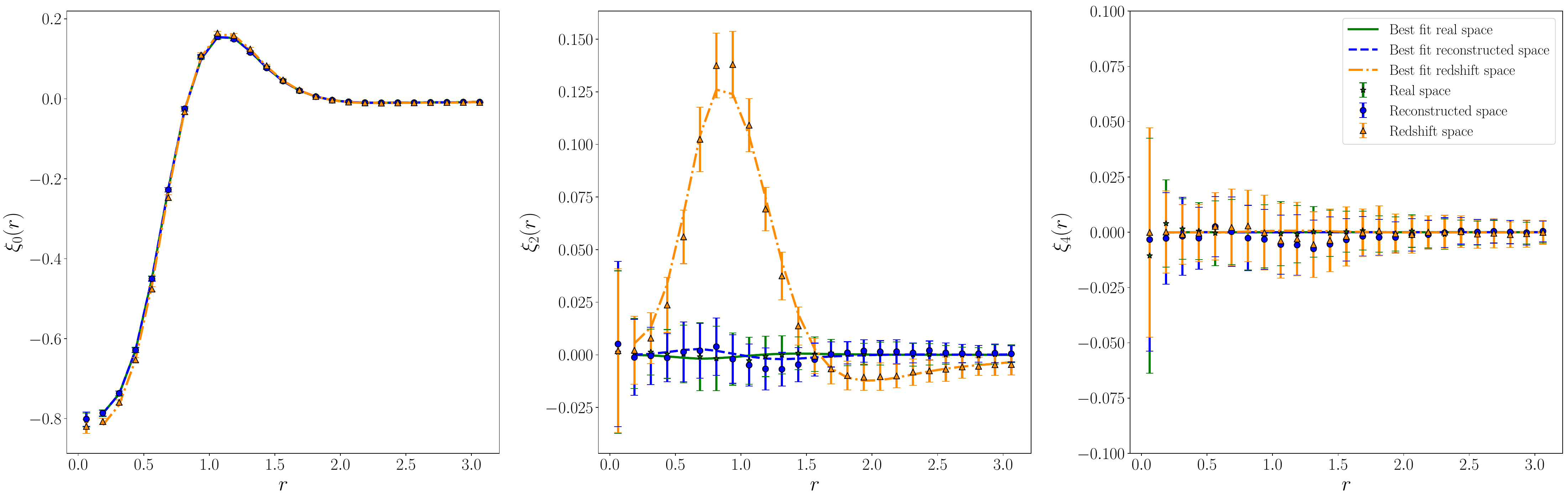}
    \caption{Monopole (left), quadrupole (center) and hexadecapole (right) of the average VGCF measured for the 100 mocks, in redshift space (orange triangles), reconstructed space (blue dots) and real space (green stars) together with the associated best-fit models (blue dashed lines for reconstructed space, orange dash dotted lines for redshift space, and green solid line for real space) estimated via the likelihood analysis presented in Section \ref{sec: likelihood}. 
    Error bars represent the standard deviation among the 100 mocks.
    In the left panel, the monopole represents the density profile of halos inside the void region, and it is quite similar in all the three cases (real, redshift and reconstructed space). In the middle panel, the quadrupole is visibly influenced by RSD in redshift space (orange), while in reconstructed space (blue) and real space (green), it is distortion-free and consistent with zero. The hexadecapole is 0 in both redshift and reconstructed space.  Here $r$ is a dimensionless quantity representing the physical separation between the center of the void and the halo, normalized by the void effective radius $R$.}
    \label{fig: multipoles bestfit rsd recon}
\end{figure*}

To sample the posterior probability distribution of all model parameters we make use of the affine-invariant Markov chain Monte Carlo (MCMC) ensemble sampler {\tt emcee} \citep{emcee}.  
The quality of the maximum-likelihood model (best fit) is assessed via evaluation of the reduced $\chi^2$ statistics
\begin{equation}
    \label{eq: chi square}
    \chi^2_\nu=- \dfrac{2}{N_\mathrm{d.o.f.}}\ln \mathcal{L}_\mathrm{eff}\left(\pmb{\Theta}\right), 
\end{equation}
where $N_\mathrm{d.o.f.}=N_\mathrm{bin}-N_\mathrm{par}$ indicates the degrees of freedom, with $N_\mathrm{bin}$ being the number of bins of the data vector and $N_\mathrm{par}$ the number of free parameters. We use 24 bins for each multipole (excluding the first bin which contains very few counts and, therefore, is very prone to shot noise) leading to a total of 72 bins for the data vector containing monopole, quadrupole, and hexadecapole.

\subsection{Covariance matrix}
\label{subsec: covariance matrix}

All analyses are carried out using a dedicated covariance matrix for each mock, estimated via jackknife resampling of its corresponding void catalog. Given that the number of available mocks is insufficient to reliably compute a full covariance matrix across realizations, we followed a standard approach adopted in the literature and apply a delete-1 jackknife strategy to estimate the covariance for each of the 100 mocks individually \citep[see e.g.][]{hamaus_2020, hamaus_2022, radinovic_2023}. In Appendix~\ref{appendix: covariances}, we show that the jackknife-derived covariance matrices are consistent with those obtained from the full mock-to-mock covariance. Based on this agreement, we adopt the jackknife covariance, $\text{Cov}[\pmb{\xi}]_{ij}$, throughout our analysis.

This methodology relies on ergodicity, which allows us to average measurements across different spatial patches to estimate the covariance matrix \citep{hamaus_2020}. In this approach, we remove one void at a time from the sample when estimating $\xi_\ell(r)$, providing a total of $N_\mathrm{v}$ jackknife samples, where $N_\mathrm{v}$ is the total number of voids in the catalog. These samples can then be used in Eq. \eqref{eq: cov jk} to calculate $\text{Cov}[\pmb{\xi}]_{ij}$, with an additional normalization factor of $(N_\mathrm{v}-1)$ to account for the statistical weight of the jackknife sample size. The elements of the covariance matrix are computed as follows
 \begin{equation}
    \label{eq: cov jk}
    \text{Cov}[\pmb{\xi}]_{ij}=\frac{N_\mathrm{V}-1}{N_\mathrm{V}}\sum_{k=1}^{N_\mathrm{V}}\left(\pmb{\xi}^{(k)}_i-\bar{\pmb{\xi}}_i\right) \left(\pmb{\xi}^{(k)}_j-\bar{\pmb{\xi}}_j \right) \,
\end{equation}
where the sum runs over the $N_\mathrm{V}$ jackknife samples, with the indices $i$ and $j$ referring to the bins of the measured VGCF multipoles.

\section{Results}
\label{sec: results}

Our analysis has two main goals. 
\begin{itemize}
    \item The first -- a pre-requisite for the second -- is to assess the performance of the reconstruction algorithm in removing RSDs. To do so, we focus on the parameter $\beta$, which quantifies the amplitude of RSD effects in the VGCF multipoles. In the ideal case of perfect RSD removal, we expect $\beta = 0$. In practice, however, uncertainties are expected to arise from imperfect dynamical reconstruction, void detection, and VGCF estimation. To isolate the effects introduced solely by reconstruction and assess their impact on the void-finding procedure, we compared the best-fit values of $\beta$  obtained from the reconstructed catalogs with those measured directly in real space, rather than with the expected value $\beta = 0$. A close match between the two $\beta$ values would indicate that the reconstruction effectively removes RSDs.

    \item The second is to evaluate whether performing the AP test in reconstructed space yields tighter constraints on the geometric distortion parameter $\varepsilon$, compared to using voids identified in redshift space. To this end, we compared the AP constraints derived from both redshift and reconstructed spaces void catalogs using the full void sample (Section~\ref{subsec: redshift vs recon space}). We then assessed the impact of excluding small voids, motivated by their greater susceptibility to noise and systematics, by repeating the analysis on size-selected subsamples in both spaces (Section~\ref{subsec: void radii}). 

\end{itemize}

Finally, to assess the robustness of the results presented in this section, we performed a series of consistency tests. These include examining the sensitivity of the method to the parameters used in the reconstruction process and evaluating its stability with respect to the choice of the covariance matrix. A detailed description of these tests and their outcomes is provided in Appendix~\ref{appendix: robustness tests}.

\subsection{Redshift vs. Reconstructed space}
\label{subsec: redshift vs recon space}

The data vector in redshift space, $\pmb{\xi}^s = \left( \xi^s_0, \xi^s_2, \xi^s_4 \right)$, consists of the three even multipoles of the VGCF estimated in redshift space. Similarly, the data vector in reconstructed space, $\pmb{\xi} = \left( \xi_0, \xi_2, \xi_4 \right)$, contains the same multipoles estimated from the VGCF in reconstructed space.

Fig.~\ref{fig: multipoles bestfit rsd recon}, shows the three measured multipoles averaged for the 100 mocks (described in Section  \ref{sec: data}) -- in redshift space (orange triangles and solid curves), reconstructed space (blue dots and solid curves) and real space (green stars and solid curves). 
Focusing on the quadrupole (central panel), we observe a positive bump near the ridge at $r \sim 1$ in redshift space. This feature signals the presence of coherent outflows and manifests as an apparent elongation of the void along the LOS. On the other hand, the quadrupole measured in reconstructed space is compatible within error bars with the one in real space, with both being consistent with zero, indicating the effective removal of RSDs. The consistency of the hexadecapole with zero further supports this conclusion.

\begin{figure}[htbp]
    \centering
    \includegraphics[width=0.9\linewidth]{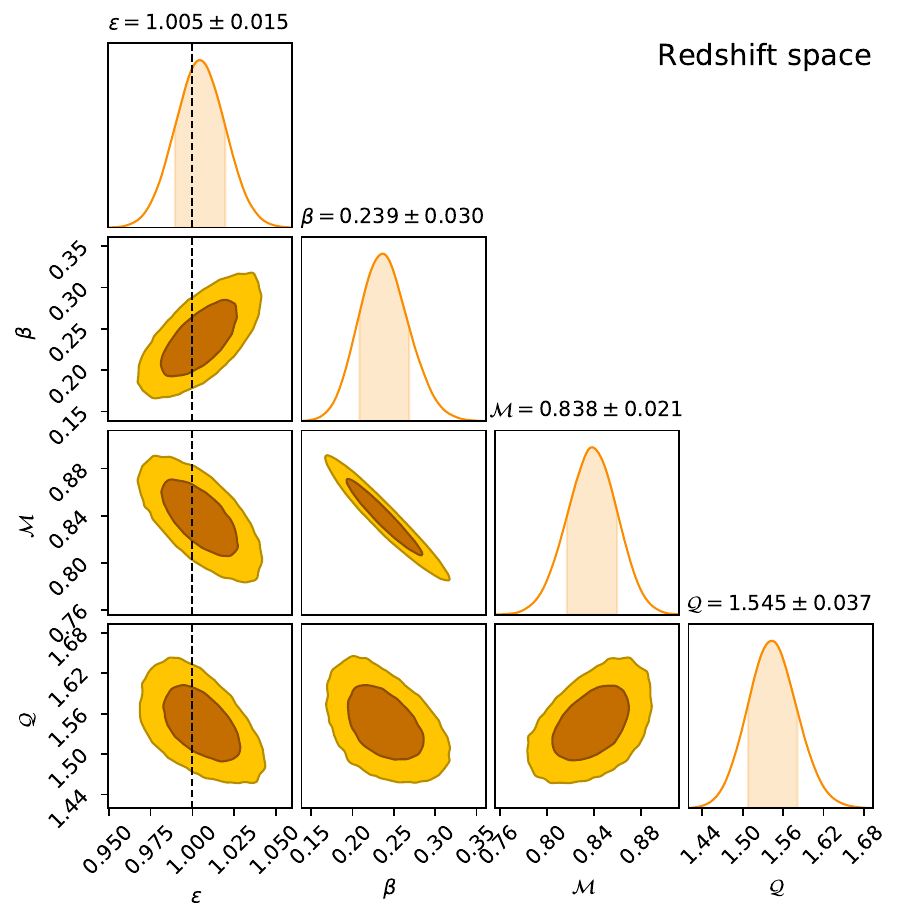}
    \caption{Posterior probability distribution of the model parameters that enter in Eq. \eqref{eq: xi model RSD MQ}, obtained via MCMC from the data vector in redshift space shown in orange in Fig. \ref{fig: multipoles bestfit rsd recon}. Dark and light-shaded areas represent $68\%$ and $95\%$ confidence regions, and dashed lines indicate fiducial values of the RSD and AP parameters $\beta$ and $\varepsilon$. The top of each column states the mean and standard deviation of the 1-dimensional marginal distributions.}
    \label{fig: chains_rsd}
\end{figure}
\begin{figure}[htbp]
    \centering
    \includegraphics[width=0.8\linewidth]{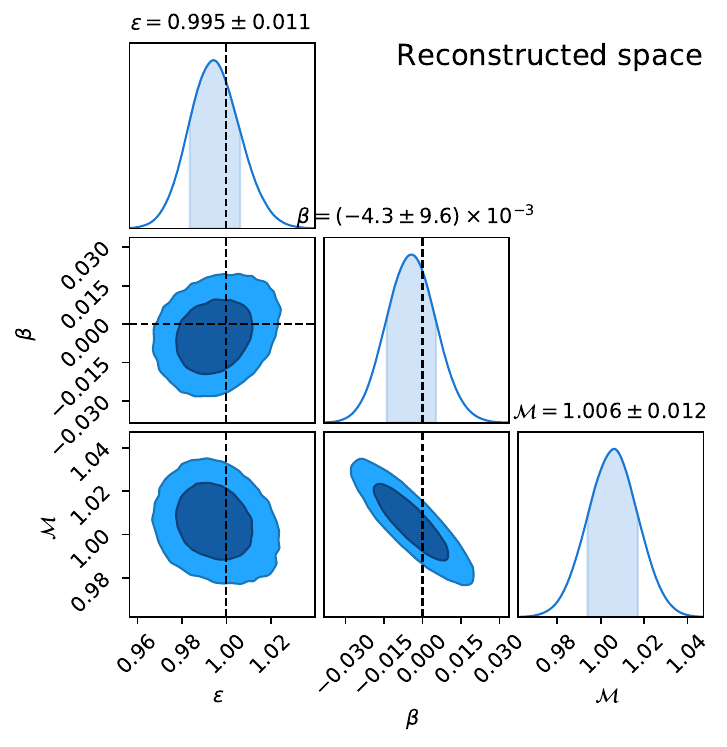}
    \caption{Posterior probability distribution of the model parameters that enter in Eq. \eqref{eq: xi model RSD MQ}, obtained via MCMC from the data in reconstructed space, shown in blue curve in Fig. \ref{fig: multipoles bestfit rsd recon}. Dark and light-shaded areas represent $68\%$ and $95\%$ confidence regions, and dashed lines indicate fiducial values of the RSD and AP parameters. The top of each column states the mean and standard deviation of the 1-dimensional marginal distributions.}
    \label{fig: chains_recon}
\end{figure}

We perform a full likelihood analysis in both redshift space and reconstructed space, as described in Section~\ref{sec: likelihood}. The multipoles of the best-fit model are shown in Fig.~\ref{fig: multipoles bestfit rsd recon}, represented by the dash-dotted purple line for redshift space and the dashed red line for reconstructed space. The corresponding reduced chi-square values are $\chi^2_\nu = 1.31$ for redshift space and $\chi^2_\nu = 1.15$ for reconstructed space, respectively. Notably, the analysis in reconstructed space, which involves one fewer free parameter than its redshift-space counterpart, yields a lower reduced chi-square value.

The posterior distributions for the model parameters in redshift space $\pmb{\Theta}=\left( \varepsilon, \beta, \mathcal{M}, \mathcal{Q}\right)$, are illustrated in Fig.~\ref{fig: chains_rsd}, while for the model parameters in reconstructed space $\pmb{\Theta}=\left( \varepsilon, \beta, \mathcal{M}\right)$, are illustrated in Fig.~\ref{fig: chains_recon}. The darker and lighter-shaded areas represent the $68\%$ and $95\%$ confidence regions. Dashed vertical lines indicate the expected values of the parameters $\beta$ and $\varepsilon$. The mean and standard deviation of the 1-dimensional marginal distributions are shown on the top of each column. \\
We first focus on the parameter that is central to this work, $\varepsilon$, whose true value is expected to be unity. The best-fit values of $\varepsilon$ are $1.005 \pm 0.015$ in redshift space, and $0.995\pm 0.011$ in reconstructed space. The relative precision of $\varepsilon$ in redshift space is $1.5\%$, with the best-fit value consistent with the expected result at the $0.3\sigma$ level. After applying reconstruction, the precision improves to $1.1\%$, while the best-fit remains consistent with unity within $0.5\sigma$. This corresponds to an improvement in statistical precision by a factor $ I \equiv \sigma_{\mathrm{redshift\ space}}/\sigma_{\mathrm{reconstructed\ space}} = 1.3, $ comparable to that achieved when applying Zel’dovich reconstruction to BAO analyses \citep[I=1.6, ][]{ross_2017}. These results demonstrate that reconstruction (when correlating voids and galaxies both in reconstructed space) provides a clear and quantitative gain in the precision of void-based AP measurements. Reconstruction therefore establishes itself as a key tool for precision void cosmology.

Moving to the other relevant parameter, $\beta$, we observe that its value in the redshift-space analysis, see Fig.~\ref{fig: chains_rsd}, corresponds to the standard RSD distortion parameter $f/b$. In contrast, in the reconstructed space, where RSDs are expected to be removed, any deviation from the expected value of zero would indicate an imperfect reconstruction.
We find that in the reconstructed-space analysis, $\beta = -0.004 \pm 0.010$, consistent with zero and confirming the effectiveness of the RSD removal. In redshift space, however, we measure $\beta = 0.239 \pm 0.030$, which shows a significant $5.7\sigma$ discrepancy compared to the expected value of $\beta = 0.408$ (Section~\ref{sec: data}). \\
Several factors may contribute to this tension. First, the analysis includes all voids identified in the mock catalogs, including smaller ones that are more susceptible to nonlinear dynamics and excluded in previous analyses. Second, systematic effects inherent to the redshift-space void identification process, particularly selection biases, may play a significant role \citep{correa_2022}. Third, the effective halo bias within voids, especially smaller ones, may deviate significantly from the one estimated from the halo-halo correlation function in the mocks  \citep{pollina_2017, verza_2022}, potentially impacting the accuracy of the expected $\beta$ value.

One of the central motivations of this work is to develop a methodology that enables the estimation of the geometric distortion parameter $\varepsilon$ independently of the systematics that affect the measurement of $\beta$ in redshift space, systematics that can drive $\beta$ far from its true value. Disentangling these contributions is particularly important because, in redshift space, the parameters $\varepsilon$ and $\beta$ are strongly correlated. This is evident in Fig.~\ref{fig: chains_rsd}, where their joint posterior displays a pronounced alignment, with a correlation coefficient of $\mathcal{R}=0.67$ estimated directly from the parameter chains. Since performing a robust AP test requires an accurate treatment of RSD, applying reconstruction methods proves highly effective: as shown in Fig.~\ref{fig: chains_recon}, the posterior of $\varepsilon$ and $\beta$ becomes nearly spherical after reconstruction -- and the correlation is reduced to $\mathcal{R}=0.23$ -- demonstrating that the degeneracy between RSD and AP effects is largely mitigated.
 \\
As for the nuisance parameters $\mathcal{M}$ and $\mathcal{Q}$, they do not carry direct cosmological significance \citep{hamaus_2022}, and there are no expected values against which they can be compared. Nevertheless, it is reassuring that in reconstructed space, the best-fit value of $\mathcal{M}$ is consistent with unity, an outcome expected if redshift-space distortions have been effectively removed. To further assess the quality of the analysis performed in reconstructed space and to perform a consistency test, we compared the results with those obtained from an analysis in real space. In this third analysis, a data vector consisting of the VGCF multipoles measured in real space is input into the same likelihood framework used for the analysis in reconstructed space. The best-fit values from the real-space analysis, $\varepsilon = 1.002 \pm 0.012$ and $\beta = -0.001 \pm 0.010$, are in excellent agreement with those obtained in reconstructed space, both in terms of central values and uncertainties (Table~\ref{tab: results eps beta}). This result shows that reconstruction delivers the best precision attainable with the given galaxy catalog, model, and void finder. 

\begin{table}[h]
    \centering
    \caption{Parameter constraints for $\varepsilon$ and $\beta$.}
    \begin{tabular}{| c  |c  |c  |c|}\hline
 & Redshift space& Recon space&Real space\\\hline
 $\varepsilon \pm \sigma_\varepsilon$& $1.005 \pm 0.015$& $0.995 \pm 0.011$&$1.002\pm 0.012$\\\hline
 
        $\beta \pm \sigma_\beta$& $0.239 \pm 0.030$& $-0.004 \pm 0.010$&$-0.001\pm 0.010$\\\hline
    \end{tabular}
    \tablefoot{Table summarizing the values of the parameters $\varepsilon$ and $\beta$, mean values with $68\%$ confidence intervals, for the analyses in redshift, reconstructed, and real space. The precision in reconstructed space is increased by $\sim 23\%$ with respect to redshift space, and approaches the real-space analysis precision.}
    \label{tab: results eps beta}
\end{table}

\subsection{Sensitivity to the void size}
\label{subsec: void radii}
In previous analyses, all voids found by the {\tt VIDE} algorithm are included in the VGCF measurements. However, small voids increase systematic errors and are often excluded; for instance, \cite{hamaus_2022} discard voids smaller than three times the mean tracer separation ($\mathrm{mps}$). Small voids can raise the fraction of spurious voids, which suppress the VGCF amplitude and its multipoles because spurious voids have a null quadrupole and weaker monopole \citep{cousinou_2019}. While spurious voids mainly affect small voids, their impact persists even with large voids \citep{cousinou_2019}. Moreover, small voids are more affected by nonlinear effects, complicating RSD removal. This work focuses on the dynamical aspect, particularly whether small voids bias the estimation of the parameter $\varepsilon$.

\begin{figure}[htbp]
    \centering
    \includegraphics[width=0.9\linewidth]{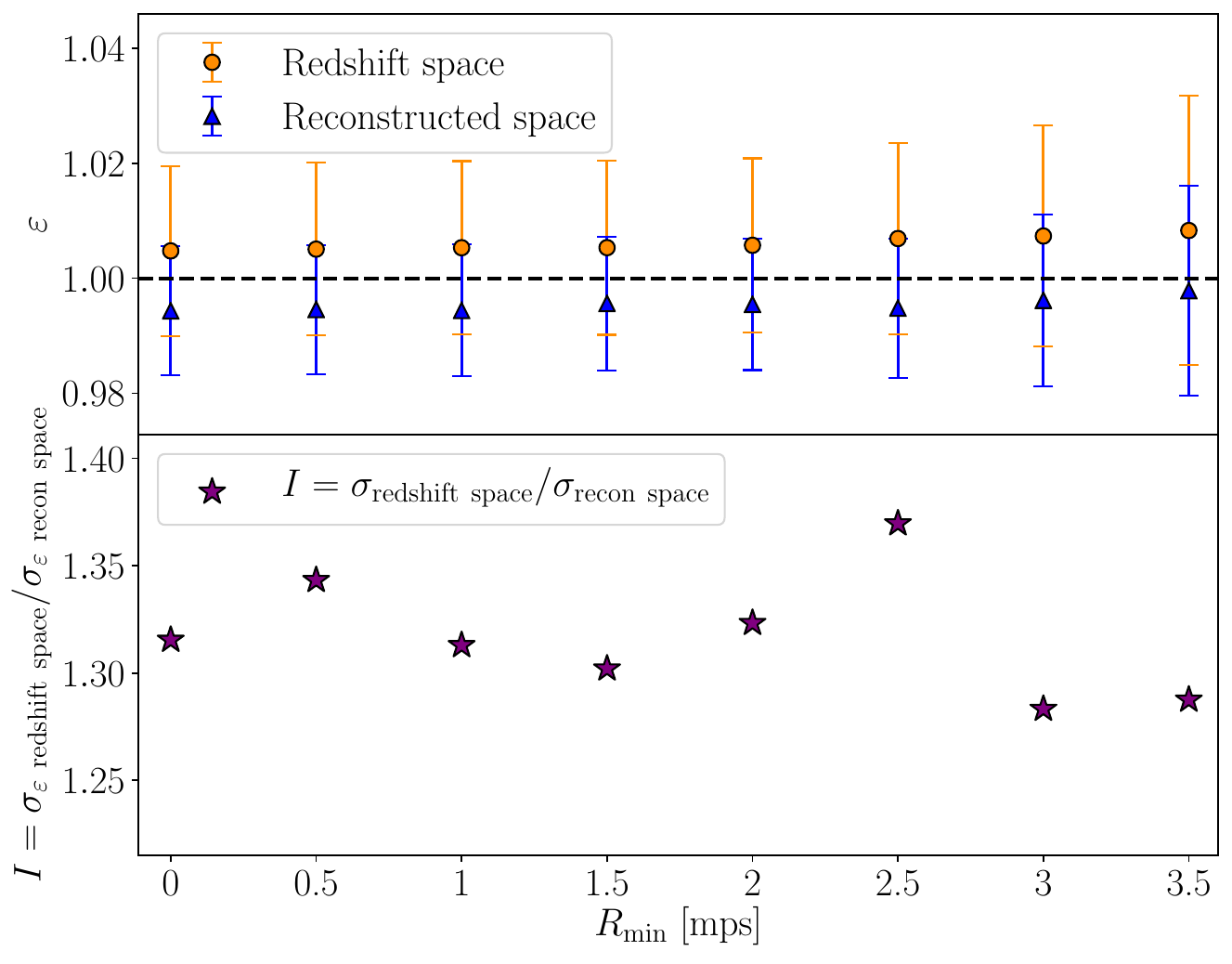}
    \caption{\textit{Top Panel:} comparison of the values of the AP parameter $\varepsilon$ with its error $\sigma_\varepsilon$ (error bars), obtained with the fitting procedure described in Section \ref{sec: likelihood}, for the analysis with tracers and voids in redshift-space (orange dots) and the analysis with tracers and voids in reconstructed space (blue triangles). Each dot represents the $\varepsilon$ value as a function of the minimum radius $R_\mathrm{min}$ for the voids in that specific subsample, expressed in mean tracer separation (mps) units, used for computing VGCF. 
\textit{Bottom Panel:} improvement factor $I$, i.e., the ratio between 
$\sigma_\varepsilon$ estimated in redshift space and reconstructed space. }
    \label{fig: epsilon rmin}
\end{figure}

To do so, we repeated the redshift-space and reconstructed-space analyses described in the previous section, progressively removing voids with radii smaller than a threshold value defined as $R_\mathrm{min} = N_\mathrm{s} \times r_\mathrm{mps}$, where $r_\mathrm{mps} = 13\ h^{-1}\mathrm{Mpc}$ denotes the mean halo separation. The results of this test are shown in Fig.~\ref{fig: epsilon rmin}, where we present the best-fit values of $\varepsilon$ (top panel) and the improvement factor $I$ (bottom panel), for redshift (orange), and reconstructed (blue) space. The error bars in the top panels correspond to the $1\sigma$ uncertainties. All quantities are shown as a function of the minimum void radius, $R_\mathrm{min}$. No significant bias in $\varepsilon$ is found regardless of void size or space. Uncertainty on $\varepsilon$ grows with $R_\mathrm{min}$ due to fewer voids, but remains consistently smaller in reconstructed space. 
The improvement factor $I$ remains stable at approximately $1.3$, with a peak at $1.35$, exhibiting a slight decrease to $1.25$ when small voids are excluded, for voids larger than $3\,R_\mathrm{min}$. 
These results corroborate the conclusion that Zel'dovich reconstruction is effective even for small-scale voids, indicating that their dynamics remain only mildly nonlinear once an appropriate Gaussian smoothing is applied to the data \citep{schuster_2022}. The fact that the improvement in precision is already detectable with only about 3,000 voids is encouraging. It suggests that applying this method to future datasets, expected to provide significantly larger void samples from denser tracer catalogs, and thus include many small voids, could yield even greater gains \citep[e.g.,][]{hamaus_2022}. By eliminating the need for a cut at small radii, reconstruction allows us to fully exploit the statistical power of these upcoming surveys. The best-fit values presented so far were obtained by analyzing the joint likelihood of the 100 mock void catalogs, with uncertainties rescaled to reflect those of a single catalog. However, when analyzing the results on a mock-by-mock basis, an interesting pattern emerges. In a few cases, the best-fit value of $\varepsilon$ significantly deviates from unity when all halos are included in the analysis. In these instances, the bias disappears once voids with sizes smaller than $3 \times \mathrm{mps}$ are excluded. Notably, this is the same cut adopted by \citet{hamaus_2022} and other analyses, to obtain unbiased results. In contrast, the reconstructed space analysis performs better even in these extreme cases, consistently yielding $\varepsilon$ values that are compatible with unity, even when small-scale voids are included. These findings demonstrate that reconstruction not only improves the precision of the $\varepsilon$ estimate but also enhances the robustness of the void-based analysis. This effect can be clearly observed in Fig.~\ref{fig: epsilon rmin mock 0} in Appendix~\ref{app: rmin single mock}, which reproduces the same analysis shown in Fig.~\ref{fig: epsilon rmin} but performed on a single mock (mock 0) instead of the full set of 100 mocks. Notably, a similar stabilizing effect of reconstruction was reported for BAO analyses \citep{sarpa_2021}, which further strengthens the case for reconstruction as a unifying framework to maximize both the precision and reliability of large-scale structure probes.

\section{Conclusions}
\label{sec: conclusions}

In this work, we present a novel approach to exploiting cosmic voids for performing the Alcock–Paczynski test. The key innovation lies in conducting the entire analysis in reconstructed space, including void detection and cross-correlation measurements, after displacing tracers from their redshift-space positions to their real-space positions using the Zel’dovich approximation. This strategy enables the inclusion of numerous small-scale voids, which are typically excluded in standard analyses, thereby increasing the sample size and, consequently, enhancing the precision of cosmological parameter estimation. \\
We evaluated the performance of this strategy using simulated datasets extracted from the Quijote simulations (see Section \ref{subsec: mock}). The analysis compared the VGCF measured in reconstructed space, where both voids and halos are placed at their reconstructed positions, with the traditional approach performed in redshift space, following methodologies established in the literature \citep[see, e.g.,][]{hamaus_2020, hamaus_2022}. The results, presented in Section~\ref{sec: results}, confirm the effectiveness of the proposed method.\\
We demonstrated that the reconstruction algorithm successfully removes RSD, as shown by the absence of a quadrupole moment in the measured VGCF. Robustness tests (Appendix~\ref{appendix: robustness tests}) show that the results are stable under $\sim 4$ \% variations in the input $\beta$, and that the optimal smoothing scale, set by the tracer density and calibrated on mocks, is $5\ h^{-1}$Mpc. The analysis is also robust to the choice of covariance matrix, whether estimated via jackknife or from the 100 mock catalogs.\\
Importantly, performing the analysis in reconstructed space leads to a $\sim 23$ \% improvement in the precision of the $\varepsilon$ estimate compared to the corresponding redshift-space analysis. The same precision is achieved when using the real space catalog, demonstrating that our methodology extracts the best information attainable for the given simulation, void finder, and VGCF model. Moreover, reconstruction significantly reduces the degeneracy between $\beta$ and $\varepsilon$, which complicates redshift-space analyses.\\
Another key advantage of the approach proposed here is that it enables the use of the full void sample, without the need to discard small voids. We have shown that including progressively smaller voids does not degrade the quality of the analysis; on the contrary, it steadily improves its precision as it allows to cleanly extract additional information. \\
This work presents the first analysis performed entirely in reconstructed space. It marks a conceptual difference from previous studies, which correlated voids identified either in redshift space \citep[e.g., ][]{hamaus_2020, hamaus_2022, verza_2024roman} or in reconstructed space \citep[e.g., ][]{nadathur_2020, woodfinden_2022, radinovic_2023} with galaxies in redshift space. The aforementioned analyses provide a broader pool of cosmological information, since they simultaneously exploit RSD and AP effects, constraining both $\beta$ and $\varepsilon$, but at the cost of increased modeling complexity. In particular, the validity of analytical RSD models around voids is still debated, and identifying voids directly in redshift space can introduce systematic effects such as void fragmentation and the violation of pair conservation between real and redshift space, as discussed in \cite{correa_2022}. In contrast, our method isolates the AP signal by performing the entire analysis in reconstructed space, where RSDs are largely removed with reconstruction. This disentangling of distortions, as shown by the reduction of the degeneracy between $\beta$ and $\varepsilon$, allows a cleaner and more precise measurement of the AP parameter $\varepsilon$, which directly probes $D_\mathrm{A}(z)H(z)$. The strength of our methodology lies in this conceptual simplification: by avoiding explicit RSD modeling and the associated uncertainties, it provides a robust framework for more precise constraint of the AP test with voids.\\ This methodology, however, also has its own sources of systematics. The reconstruction technique adopted here is sensitive to the choice of the fiducial cosmology and to the physical scale used to smooth the mass-density field traced by the halos. An incorrect choice of these quantities introduces systematic errors in the reconstructed field and, ultimately, in the void identification and analysis. To assess the magnitude of these potential effects, we performed a series of tests using a suite of mock halo catalogs (see Appendix \ref{appendix: robustness tests}). These tests show that, while the reconstruction is robust to the choice of the background cosmological model, provided its parameters lie within a plausible range consistent with current observations, it is more sensitive to the choice of the smoothing filter. The latter must therefore be optimized using dedicated, realistic mock catalogs that reproduce the correct number density of tracers. This is particularly important when analyzing real data, where the spatial number density of tracers varies across the survey volume and systematic errors must remain smaller than the already small statistical uncertainties. Importantly, one may also note that the parameter $\beta$, being effectively treated as a nuisance parameter in the post-reconstruction analysis, should naturally absorb any residual reconstruction errors, further mitigating their impact. \\ In addition to reconstruction-related uncertainties, it is also worth acknowledging that further systematics may arise from the choice of the void-finding algorithm itself. In particular, ZOBOV-based void catalogs (and generically, for some of the mentioned effects, any void catalog from survey data), typically used in observational analyses, may be affected by survey geometry, tracer selection effects, the number density of the sample, and the watershed segmentation procedure, all of which can impact void properties and, potentially, the AP signal extracted from the VGCF. Since our assessment is conducted on cubic simulation boxes, these algorithm-dependent effects are not entirely captured here and must be tested explicitly under realistic observational conditions. More generally, it is worth distinguishing between the systematics that have been explicitly studied in this work, namely those associated with reconstruction, and those that remain to be explored, including void-finder dependent effects. \\
However, even the systematics already characterized here will need to be reassessed in a more realistic setting, where the AP effect is properly accounted for. The study of these systematics in the context of the era of big data for voids will be the subject of future work, where we plan to investigate them under controlled conditions. Specifically, we will employ lightcone mocks with varying cosmologies, differing from that of the simulations to introduce a realistic AP effect, in order to rigorously quantify and control systematic biases.\\
In addition, the sensitivity of our analysis to the numerical resolution of the simulations and to the discrete sampling of galaxies represents another important aspect that deserves dedicated investigation in future work, ideally relying on dense light-cones reproducing surveys’ features. 
By outlining this roadmap, our work establishes a solid foundation for the robust application of reconstructed void-based AP analyses to forthcoming large-scale structure surveys, thereby maximizing the cosmological information extractable from future data.\\ 
This work will be especially valuable for enhancing the statistical power of VGCF analyses with the upcoming spectroscopic galaxy catalogs from DESI \citep{DESI_Collaboration_2022}, Euclid \citep{mellier_2024_euclid}, Roman \citep{spergel_2015, dore_2019}, and SPHEREx \citep{dore_2018}. Before applying the method to observational data, further testing will be required to refine its performance in the presence of AP distortions in light-cone geometry, improve the modeling by accounting for potential systematic biases introduced by the void-finding algorithm when performing the AP test \citep[see][]{radinovic_2024}, and incorporate survey-specific observational biases and selection effects. 
Another important step will be to investigate how different reconstruction methods impact void properties and cosmological constraints within this framework, as partly explored for example in \citet{maragliano_2025}. 

In particular, an interesting direction for future work is to analyze the different effects of reconstruction on distinct void populations, such as void-in-void and void-in-cloud types, which may exhibit different sensitivities to the reconstruction process. More generally, beyond reconstruction alone, it is important to investigate how different void classes contribute to the cosmological signal, as their distinct environments and dynamical properties may encode complementary information.\\
The novel methodology presented in this paper strengthens the role of cosmic voids for precision cosmology by improving parameter constraints from the Alcock–Paczynski test through reconstruction. It lays a solid foundation for the VGCF and void-based analyses to maximize their impact in the extraction of constraints from upcoming galaxy surveys.

\begin{acknowledgements}
GD wishes to express her gratitude to Mar Pérez Sar, Sofia Contarini, Carlos Correa, Simone Sartori, Giovanni Verza, Massimo Guidi, Alfonso Veropalumbo, Federico Marulli, and Lauro Moscardini for the valuable discussions. AP acknowledges support from the European Research Council (ERC) under the European Union's Horizon programme (COSMOBEST ERC funded project, grant agreement 101078174). GD and AP acknowledge support from the french government under the France 2030 investment plan, as part of the Initiative d’Excellence d’Aix- Marseille Université - A*MIDEX AMX-22-CEI-03. GD acknowledges INFN Genova for providing the computational resources and support essential to this work. MA acknowledges support from the Agence Nationale de la Recherche of the French government through the program ANR-21-CE31-0016-03. HMC acknowledges support from the Institut Universitaire de France.
\end{acknowledgements}

\bibliographystyle{aa} 
\bibliography{BibliographyGD}

\appendix

\section{Robustness tests}
\label{appendix: robustness tests}

In this appendix, we present a series of robustness tests aimed at evaluating the stability and reliability of the VGCF analysis. 
First, we examine how changes in the input parameters of the reconstruction algorithm affect the VGCF output and the estimated parameters $\varepsilon$ and $\beta$. 
Subsequently, we assess the robustness of the likelihood analysis by comparing the jackknife covariance matrix used in our analysis with that estimated from the mock catalogs. To perform these tests, we used a single mock halo catalog, as our goal was to assess the relative impact of the analysis choices described above, rather than to estimate absolute uncertainties. Since such relative comparisons are, to first approximation, insensitive to sample variance, we consider the results representative of what would be obtained using the full set of 100 mocks.

\subsection{Sensitivity to the choice of parameters that regulate the reconstruction}
\label{ch paper1 subsec: test reconstruction}

The reconstruction algorithm described in Section \ref{subsec: reconstruction} depends on several input parameters, with the most important being the smoothing filter radius applied to the input density field, $R_\mathrm{s}$, and the fiducial value of $f$,  assuming a fixed tracer bias. In this section, we investigate how variations in these parameters affect the results of the VGCF analysis.

To this end, we analyzed the VGCF measured after performing reconstructions with varying values of $R_\mathrm{s}$ and $f$. For each reconstruction, a new void catalog is extracted from the reconstructed halo fields, and the VGCF is measured using the correct cosmology to avoid introducing AP distortions. Each test consists of two steps. First, we focused on the quadrupole moment and perform a $\chi^2$ test to assess its consistency with the null  hypothesis of zero signal. Next, we carried out a full statistical inference analysis, similar to that presented in Section~\ref{sec: results}, to estimate the best-fit values of $\varepsilon$ and $\beta$.

\subsubsection{Sensitivity to the smoothing scale}
\label{subsubsec: sensitivity smoothing scale}

The reconstruction algorithm requires smoothing the input density field with a Gaussian filter. The optimal choice of the smoothing radius $R_\mathrm{s}$ depends primarily on the sample’s mean tracer number density, intrinsic clustering, and potentially on the specific cosmic structures and clustering analysis considered, so it must be determined on a case-by-case basis. $R_\mathrm{s}$ should be large enough to suppress strong nonlinear effects and ensure that the Zel’dovich approximation holds, but not so large that it removes genuine power and systematically underestimates the amplitude of peculiar velocities.\\
To assess the sensitivity to the choice of $R_\mathrm{s}$, we performed a test similar to the one employed by \cite{nadathur2019zeldovich} to optimize $R_\mathrm{s}$ in their reconstruction. We ran several reconstructions using the correct value of $f$ and assuming the correct cosmological model, while varying $R_\mathrm{s}$ in the range $[0,15]h^{-1}$Mpc, which includes the typical nonlinear scale around $10\ h^{-1}$Mpc. After each reconstructions, voids were identified using {\tt VIDE} and the VGCF was evaluated with the DP estimator. \\
To assess the quality of the RSD removal, we evaluated the $\chi^2$ difference between the measured quadrupole and the null signal using the jackknife covariance matrix as follows
\begin{equation}
\chi^2 = \sum_{i,j} \left[ \xi_2^{\mathrm{data}}(r_i) - \xi_2^{\mathrm{th}}(r_i) \right] \, \text{Cov}^{-1}_{ij} \, \left[ \xi_2^{\mathrm{data}}(r_j) - \xi_2^{\mathrm{th}}(r_j) \right]\ ,
\end{equation}
where $\xi_2^{\mathrm{th}}=0$ represents the expected quadrupole in real space.

\begin{figure}[htbp]
    \centering
    \includegraphics[width=0.9\linewidth]{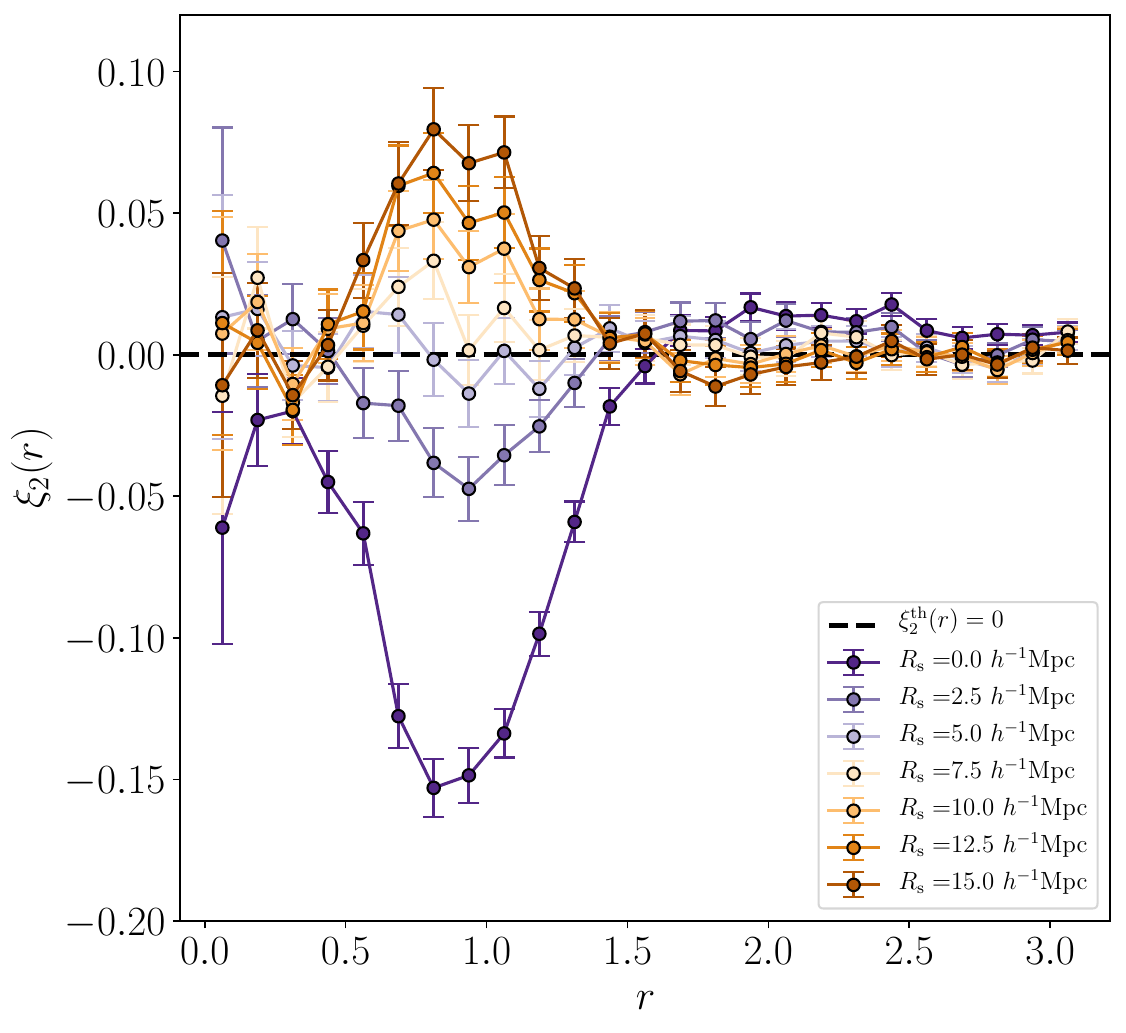}
    \caption{Quadrupole of the VGCF computed with reconstructed data, for different choices of the fixed smoothing scale $R_\mathrm{s}$ used in the reconstruction procedure. Error bars are computed as the diagonal of the covariance matrix. The choice $R_s=5.0\ h^{-1}\mathrm{Mpc}$ results in an isotropic correlation function, corresponding in a measured quadrupole matching with the dashed black line, which is the expected zero quadrupole $\xi_2^{\mathrm{th}}=0$. 
    Residual anisotropies are seen for other values of $R_\mathrm{s}$. The quadrupole moment, and consequently the effect of $R_\mathrm{s}$, is accentuated near the edge of the void due to the presence of nonlinearities caused by a higher density contrast.} 
    \label{fig: quadrupole_rs}
\end{figure}

Figure~\ref{fig: quadrupole_rs} illustrates the results of the test. The colored dots represent the VGCF measured after the reconstructions performed with different smoothing radii, as indicated in the plot. Error bars correspond to the diagonal elements of the covariance matrix. Only the choice $R_\mathrm{s} = 5 \  h^{-1}\mathrm{Mpc}$ produces a quadrupole consistent with zero. All other choices produce either a positive quadrupole (indicating void shapes elongated along the LOS) or a negative quadrupole (indicating void shapes compressed along the LOS).
 
When the smoothing scale exceeds this optimal value, large-scale modes are overly suppressed, causing an underestimation of the peculiar velocity amplitude and resulting in a residual outflow that elongates void shapes along the LOS. Conversely, using a smoothing scale that is too small leads to an over-correction of the void outflow, producing a spurious compression of their shapes along the LOS \citep{sarpa_2022}.

The values of the reduced $\chi^2$ obtained in all the explored cases listed in Table \ref{tab: rs chi square}, quantitatively confirm the visual impression of Fig.~\ref{fig: quadrupole_rs}, indicating that $R_\mathrm{s}=5\  h^{-1}\mathrm{Mpc}$ is indeed the best choice for our void-halo correlation study.

\begin{table}[h]
    \centering
    \caption{$\chi^2$ values.}
    \begin{tabular}{| c || c | c | c | c | c | c | }
    \hline
        $R_\mathrm{s} \ [ h^{-1}\ \mathrm{Mpc} ]$ & 2.5 & 5.0 & 7.5 & 10.0 & 12.5 & 15.0 \\
        \hline
        $\chi^2$ & 2.76 & 0.90 & 1.37 & 2.33 & 4.13 & 5.94 \\
        \hline
    \end{tabular}
    \tablefoot{Table of the $\chi^2$ values of the measured quadrupole $\xi_2$ fitted with the theoretically expected zero quadrupole $\xi_2^\mathrm{th}$ against different smoothing scales $R_\mathrm{s}$.}
    \label{tab: rs chi square}
\end{table}

\begin{figure}[htbp]
    \centering
    \includegraphics[width=0.9\linewidth]{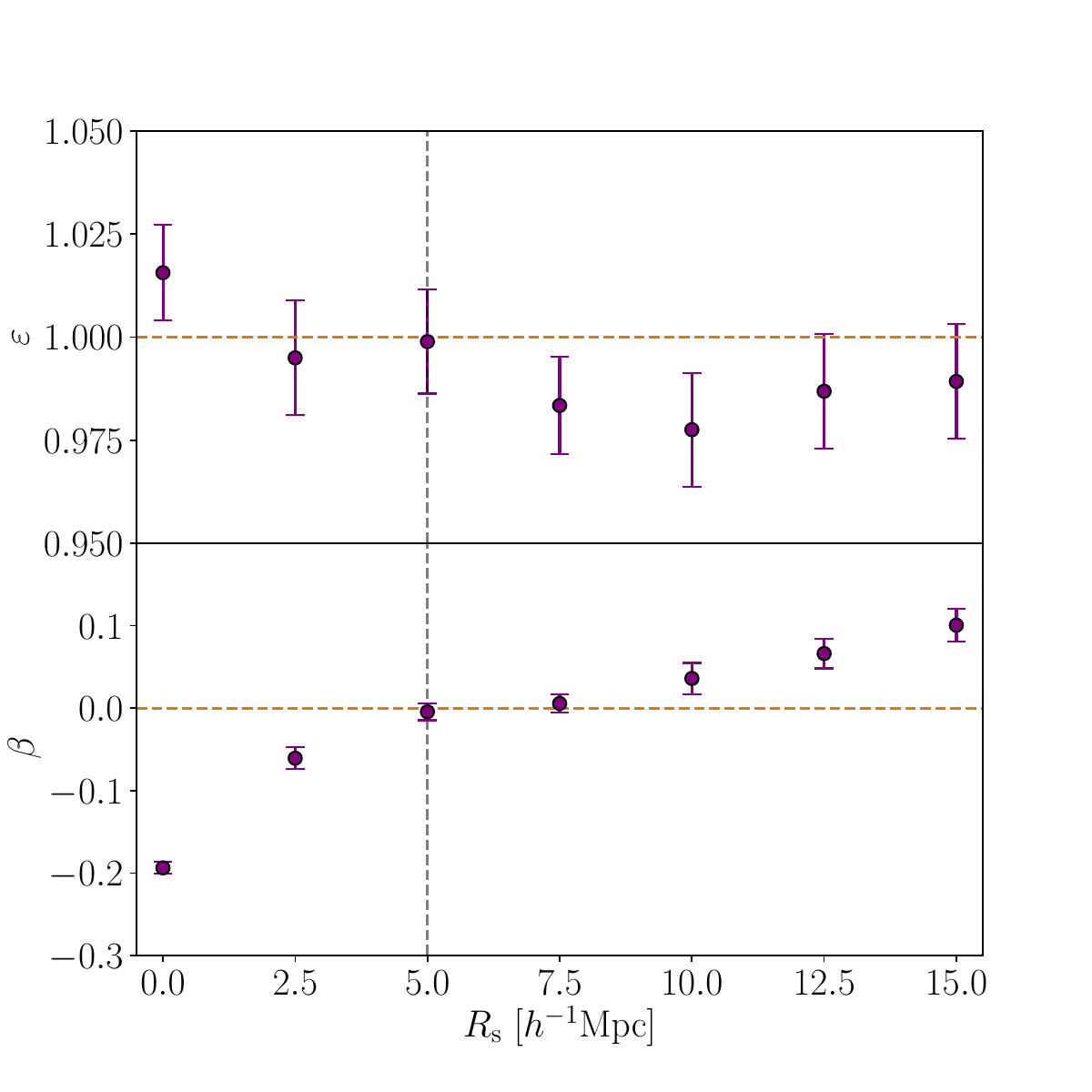}
    \caption{Estimated values of the parameters $\varepsilon$ (upper panel) and $\beta$ (lower panel) for different analyses performed in reconstructed space, each using data reconstructed with a different smoothing scale $R_\mathrm{s}$. The purple dots represent the estimated parameter values for each analysis at a specific $R_\mathrm{s}$, while the black dashed line indicates the expected values, $\varepsilon=1$ and $\beta=0$.}
    \label{fig: test rs epsilon e beta}
\end{figure}

The best-fit values of $\varepsilon$ and $\beta$ are shown in Fig.~\ref{fig: test rs epsilon e beta} as a function of $R_\mathrm{s}$ and compared to the expected values (horizontal dashed lines). For values of $R_\mathrm{s}$ smaller than the optimal one, the reconstruction yields unphysical negative values of $\beta$ reflecting the spurious negative quadrupole caused by over-correction (bottom panel). Conversely, for $R_\mathrm{s}$ values larger than the optimal scale, the quadrupole becomes positive due to residual outflows, and the best-fit analysis returns a positive $\beta$ to match the signal. On the contrary, $\varepsilon$ (top panel) shows a much weaker sensitivity to the choice of smoothing scale. Its best-fit value remains consistent with the expected value within approximately  $\sim 1 \sigma$  across the full range of $R_\mathrm{s}$ values considered.

\subsubsection{Sensitivity to the input $\beta$ parameter}
\label{ch paper1 subsubsec: sensitivity fiducial cosmology}

To assess the sensitivity of the reconstruction to the choice of the input $\beta$ value, we performed a series of reconstructions using different values of $\beta$, while keeping the cosmological model fixed to the correct one for converting redshifts into distances. This approach implicitly treats $\beta$ as a free parameter, ignoring the specific relation  between $f$ and $\Omega_m$ that holds in models like $\Lambda$CDM. \\
To define a reasonable range for varying $\beta$ in this test, we first fixed the linear bias to the value estimated from the halo 2PCF (Appendix~\ref{appendix: bias}).
Then, using the $5\sigma$ uncertainty range, with each step corresponding to $1\,\sigma$, from the Planck CMB analysis \citep{Planck2020}, we derived the corresponding range of $f$ values via the $\Lambda$CDM relation $f=\Omega_\mathrm{m}^{0.55}$. The resulting $f$ range, $[0.728,0.792]$, is centered on the true simulation value $f^\mathrm{true}=0.763$, corresponding to $\beta$ values in the range $[0.389,0.424]$, with $\beta^\mathrm{true}=0.408$.
For each value, we performed both the reconstruction and void-finding steps, then computed the VGCF multipoles using halos and voids identified in reconstructed space. Uncertainties were estimated using jackknife covariance matrices for each VGCF measurement.

\begin{figure}[htbp]
    \centering
    \includegraphics[width=1\linewidth]{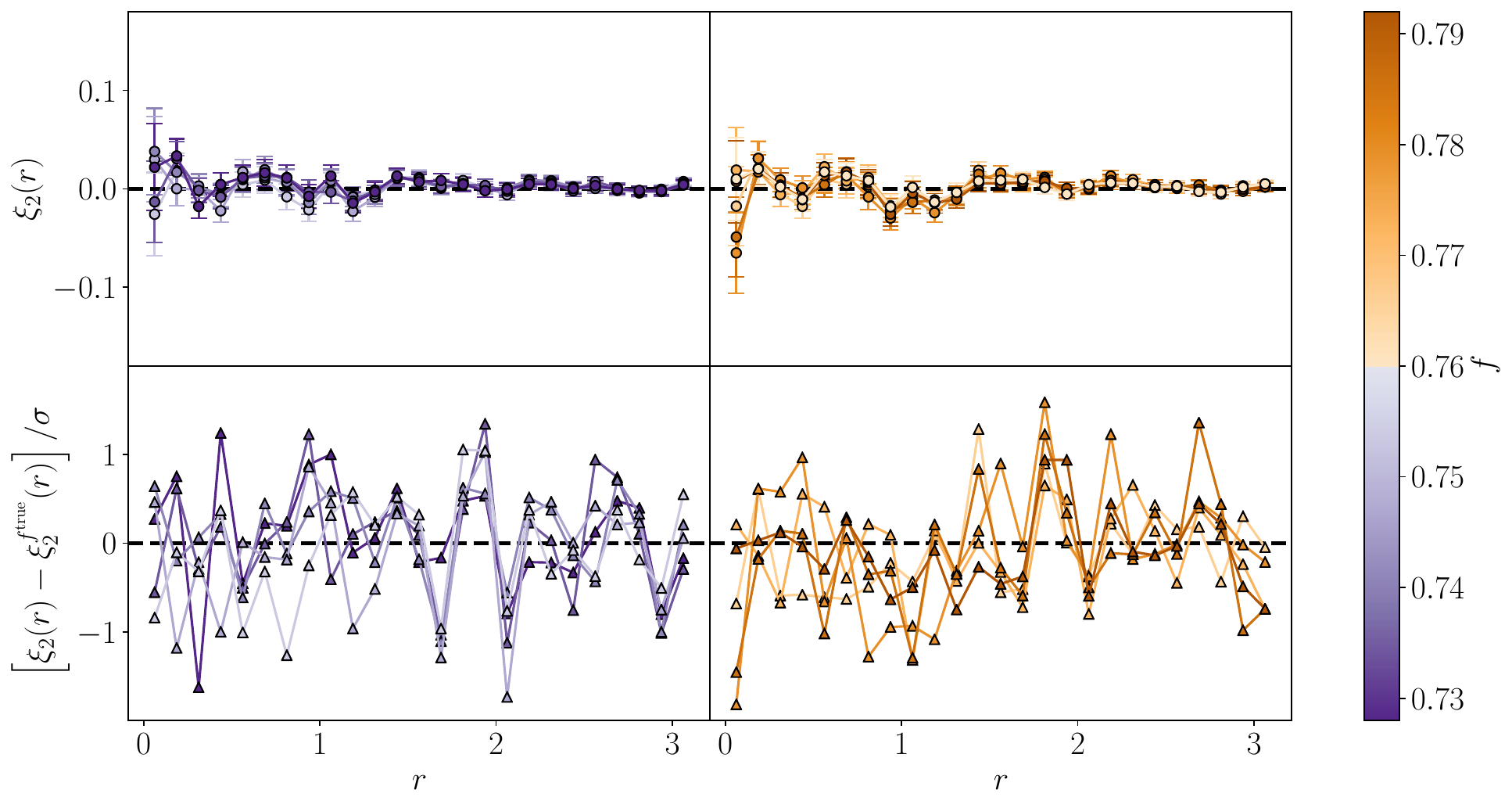}
\caption{\textit{Top Panel}: Quadrupole $\xi_2$ of the stacked void-galaxy cross-correlation function computed with voids and halos in reconstructed space, where reconstruction is computed with different values of  $f$, separating the values of $f< f^\mathrm{true}$ (left) and $f> f^\mathrm{true}$ (right). \textit{Bottom Panel}: Residuals between the quadrupoles of the top panel and the reference quadrupole computed with the fiducial cosmology of the simulation $\xi_2^{f=\text{true}}$, separating the values of $f< f^\mathrm{true}$ (left) and $f> f^\mathrm{true} $ (right). The color bar indicates the different values of $f$. Error bars are computed with the diagonals of the jackknife covariance matrices.}
    \label{fig: xi2 residuals Test Om}
\end{figure}

The top panels of Fig.~\ref{fig: xi2 residuals Test Om} show the quadrupole moments computed using different $f$ values in the reconstruction. Different colors, as indicated in the vertical bar, correspond to various choices for $f$. To avoid overcrowding the plot, we separate the reconstructions performed using $f< f^\mathrm{true} $ (left) from those performed with $f> f^\mathrm{true}$ (right).

\begin{figure}[htbp]
    \centering
    \includegraphics[width=0.9\linewidth]{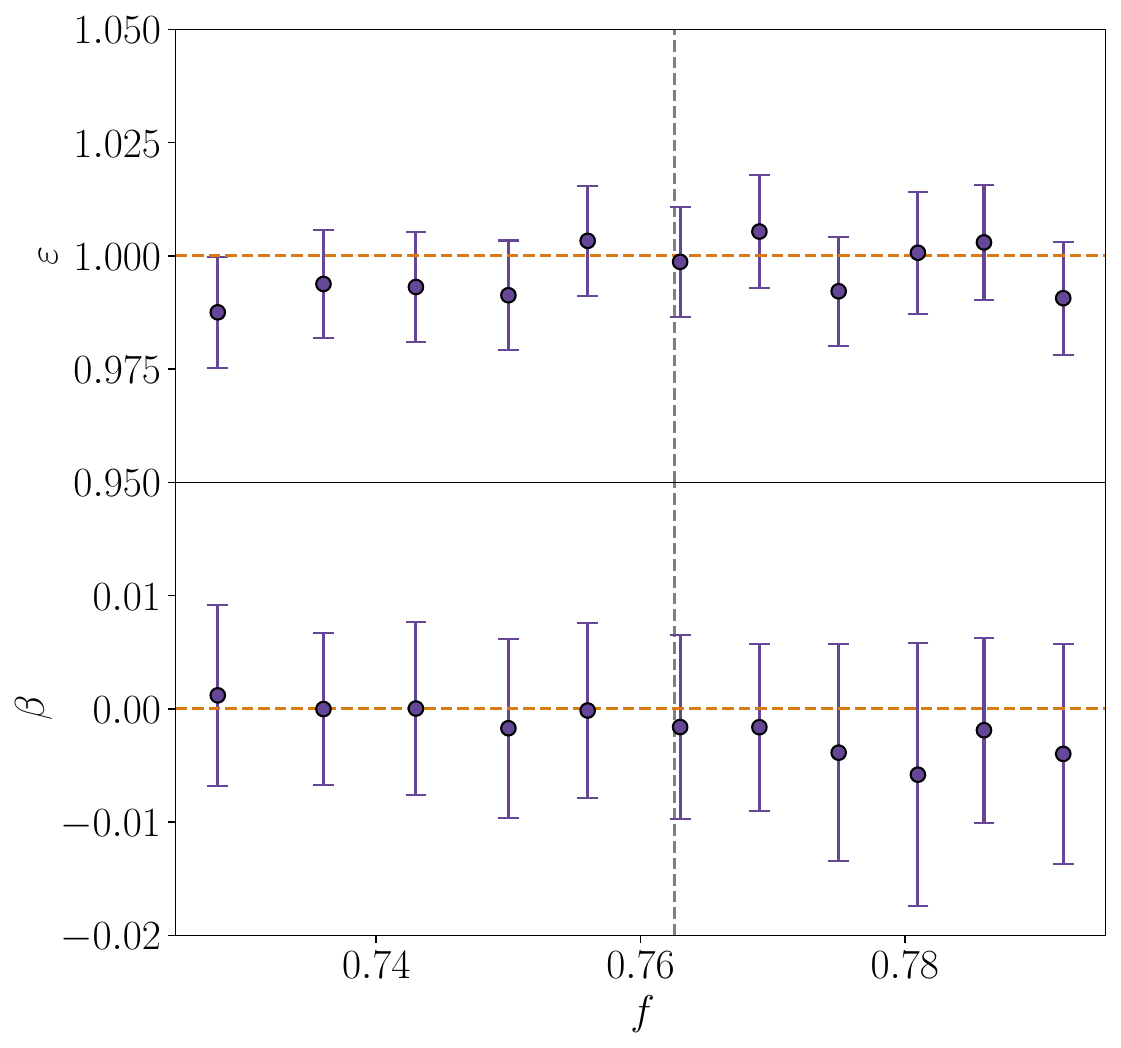}
    \caption{Estimated values of the parameters $\varepsilon$ (upper panel) and $\beta$ (lower panel) for different analyses performed in reconstructed space, each using data reconstructed with different values of the input parameter $f$. The purple dots represent the estimated parameter values for each analysis at a specific $f$, while the orange dashed lines indicate the expected values, $\varepsilon=1$ and $\beta=0$. The black vertical dashed line indicates the true value of $f$.}
    \label{fig: test f epsilon beta}
\end{figure}

We again used the quadrupole as a proxy for assessing reconstruction quality, where significant deviations from zero indicate reconstruction inaccuracies. To aid the visual interpretation, we plot the residuals (divided by the error) relative to the reference quadrupole in the bottom panels. Visual inspection reveals no significant systematic departures from the reference reconstruction and no discernible trend with increasing or decreasing $f$. The estimated reduced $\chi^2$ values confirm the visual inspection, with most values close to unity and none reaching 2. Finally, in Fig.~\ref{fig: test f epsilon beta}, we present the best-fit values of $\varepsilon$ and the output $\beta$ as a function of the input $f$. The results demonstrate that both parameters are largely insensitive to the choice of input $f$, remaining consistent with their expected values across the tested range.\\
We conclude that the results of the VGCF analysis are insensitive to the cosmology assumed in the reconstruction when the $f$, and consequently $\beta$, parameter is varied by approximately $8\%$, which is consistent with the current uncertainty in relevant cosmological parameters.

\subsection{Sensitivity to the covariance matrices}
\label{appendix: covariances}

\begin{figure}[htbp]
\centering
\includegraphics[width=0.9\linewidth]{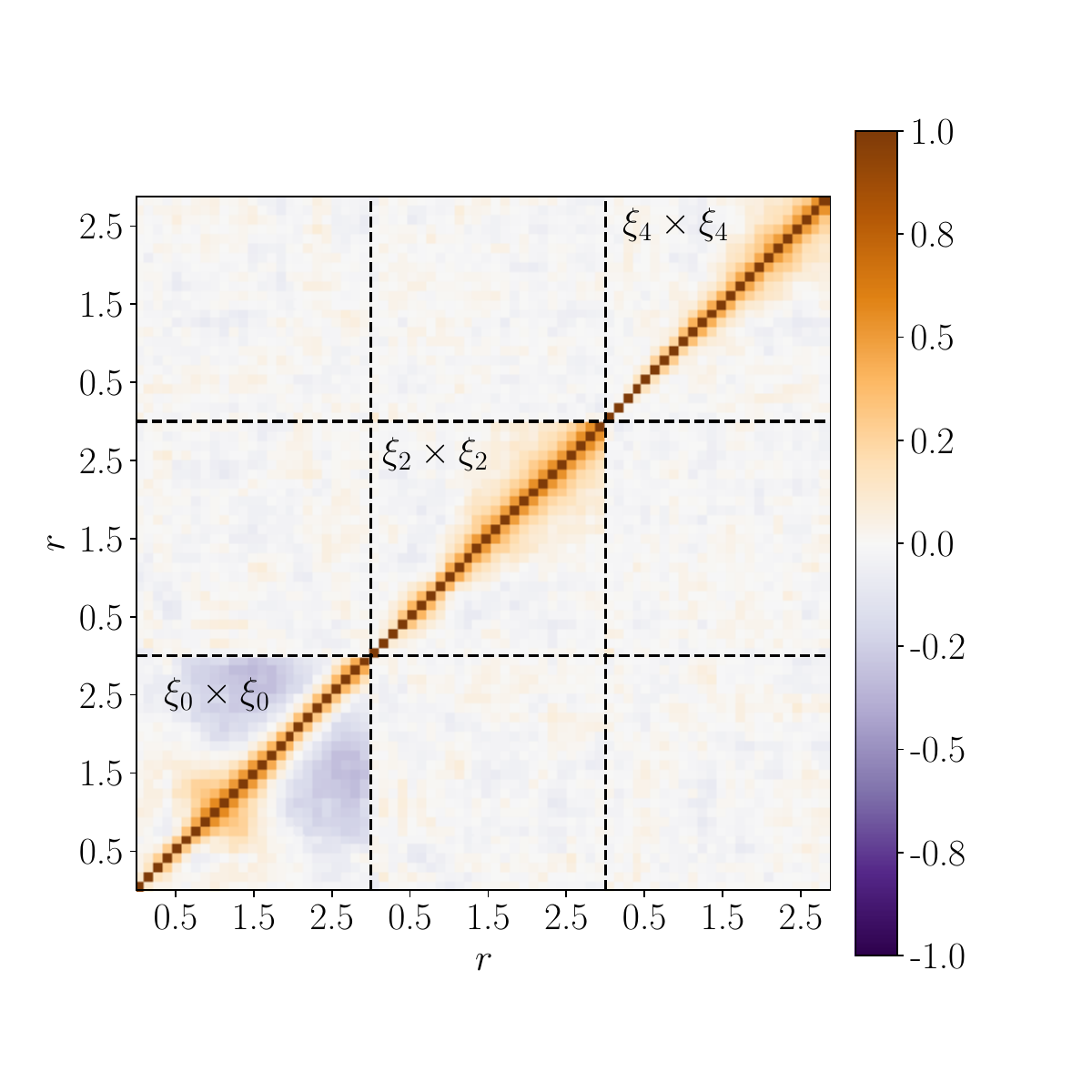}

\caption{Normalized covariance $\text{Corr}[\pmb{\xi}]_{ij}$ obtained from the jackknife resampling. }
\label{fig: cov jk}
\end{figure}

\begin{figure}[htbp]
\centering
\includegraphics[width=0.9\linewidth]{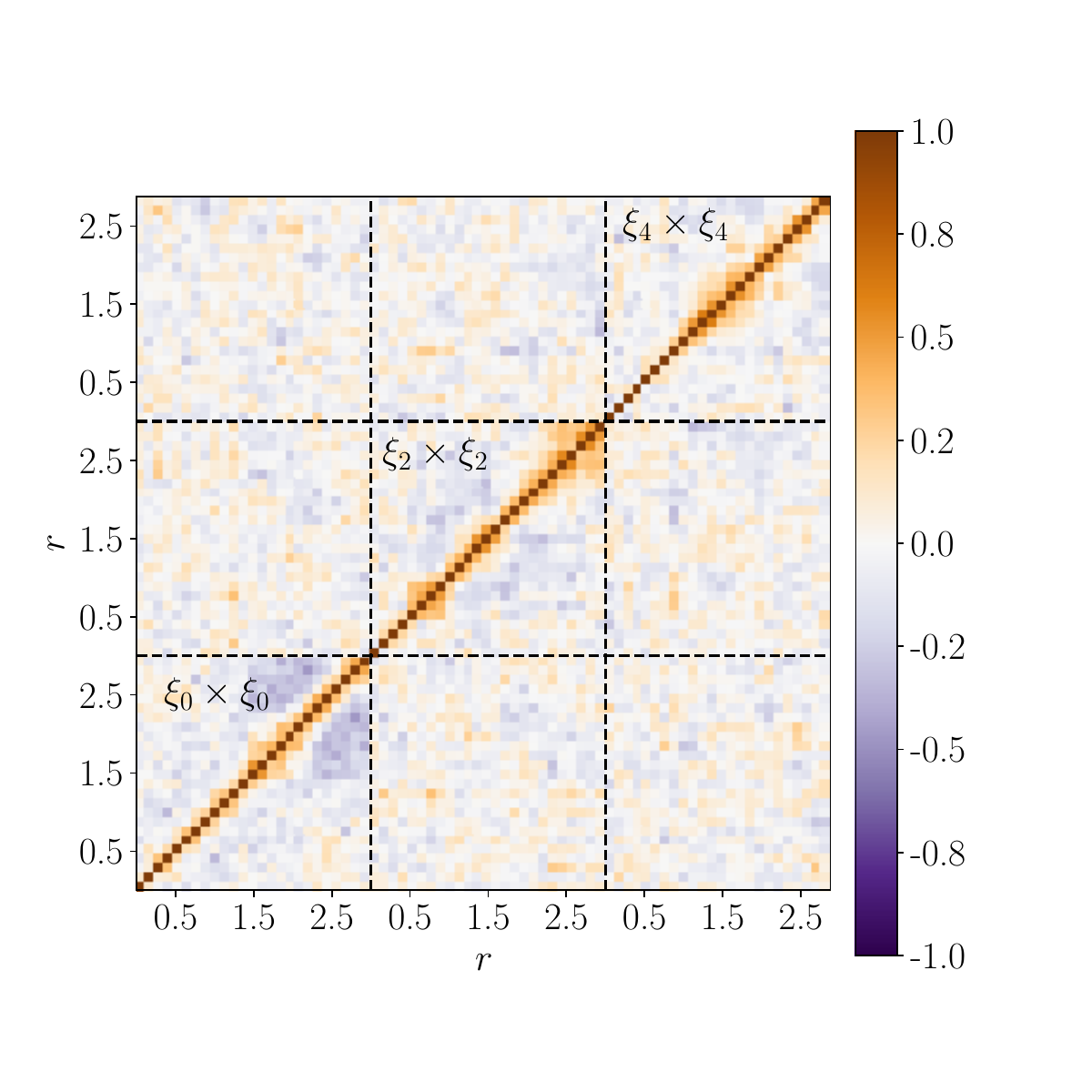}

\caption{Normalized covariance matrix $\text{Corr}[\pmb{\xi}]_{ij}$ obtained from the 100 mocks. }
\label{fig: cov mocks}
\end{figure}

\begin{figure}[htbp]
\centering
\includegraphics[width=0.9\linewidth]{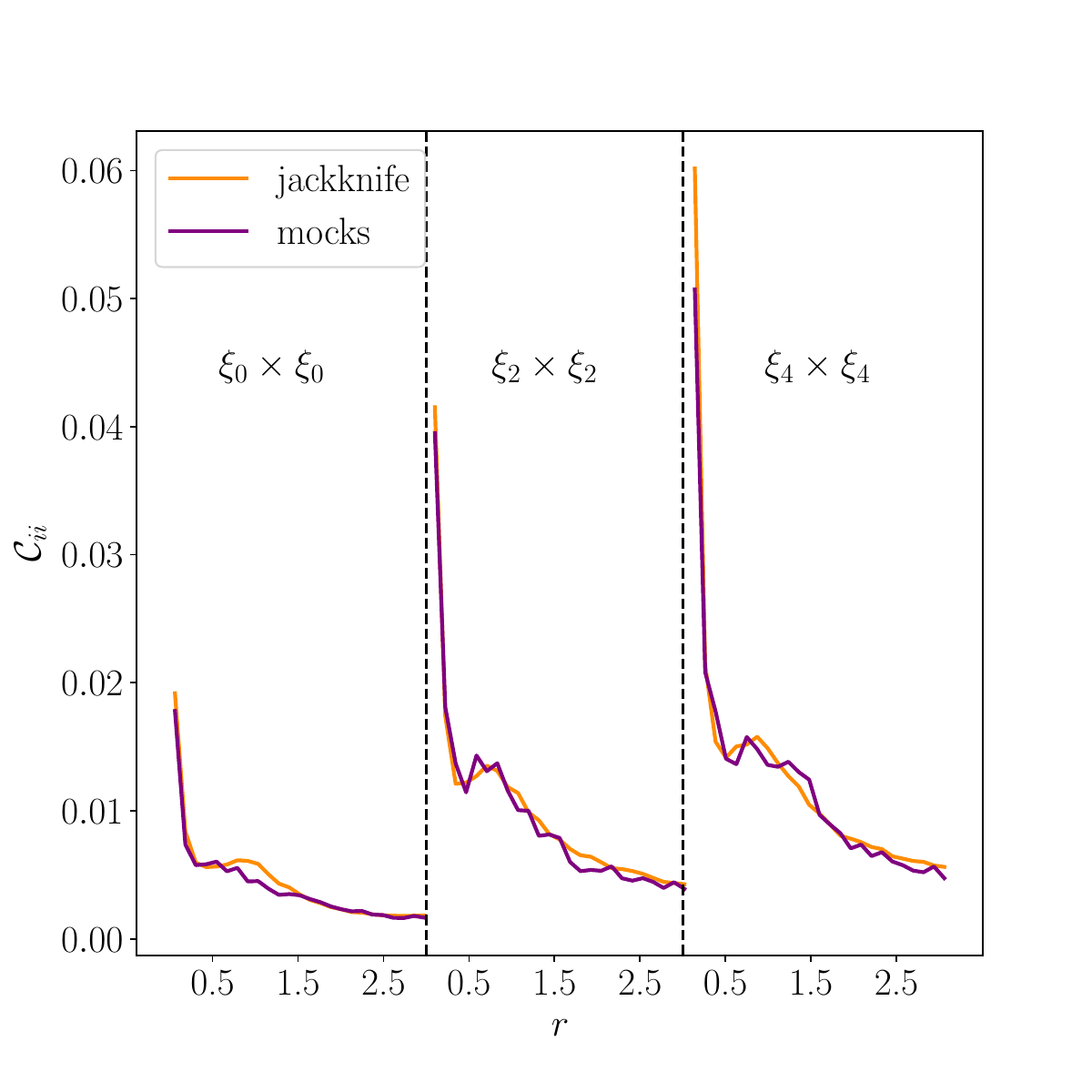}
\caption{Comparison of the elements along the diagonal, $\mathcal{C}_{ii}$ of the two covariance matrices, jackknife resampling (orange) and mock covariance (purple). Each panel of the figure represents the diagonal of the $i-$th multipole as illustrated with the text.  }
\label{fig: diagonal cov mocks}
\end{figure}

In the analyses presented in Section~\ref{sec: results}, we used a covariance matrix estimated through jackknife resampling. This approach is consistent with standard practice in the literature and is particularly advantageous in situations where mock catalogs are unavailable. Here, we consider an alternative method in which the covariance matrix is computed directly from the VGCF measurements obtained from the 100 mock catalogs. Figs.~\ref{fig: cov jk} and \ref{fig: cov mocks} show the normalized covariance matrices, defined as
\begin{equation}
   \text{Corr}[\pmb{\xi}]_{ij}=\frac{\text{Cov}[\pmb{\xi}]_{ij}}{\sqrt{\text{Cov}[\pmb{\xi}]_{ii}\text{Cov}[\pmb{\xi}]_{jj}}}\ ,
\end{equation}
computed with the two methods. The three sub matrices along the diagonal refer to each multipole. Cross terms are off-diagonal. They appear qualitatively similar. The off-diagonal elements of the covariance matrix estimated from the mocks look noisier than in the jackknife case, reflecting the limited number of catalogs available (100) compared to the number of entries in the $72\times 72$ matrix (24 entries for each multipole) \citep{hartlap_2006}.
Nonetheless, the diagonal elements of the two matrices are quite similar, as shown in Fig.~\ref{fig: diagonal cov mocks}.

\begin{figure}[htbp]
\centering
\includegraphics[width=0.8\linewidth]{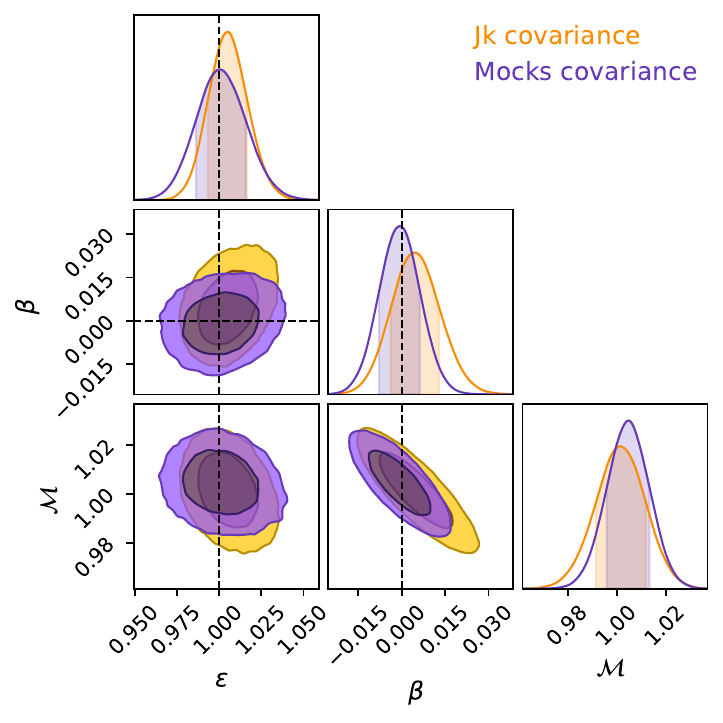}

\caption{Posterior probability distribution of the model parameters computed with the likelihood analysis presented in Section~\ref{sec: likelihood}, from the data of the single mock (mock 0), obtained using two different covariance matrices. Orange: jackknife resampling covariance matrix, and purple: mock covariance matrix corrected with factors proposed by \citep{hartlap_2006} and \citep{percival_2014}. Dark and light-shaded areas represent $68\%$ and $95\%$ confidence regions, and dashed lines indicate fiducial values of the RSD and AP parameters $\beta$ and $\varepsilon$.}
\label{fig: chains covariance comparison}
\end{figure}

To assess the quantitative impact of the covariance matrix choice, we performed the VGCF analysis in the reconstructed space using both matrices.
To correct for known biases arising from the use of a finite number of mocks, we applied the corrections proposed by \citep{hartlap_2006} and \citep{percival_2014} to the numerically estimated covariance matrix.
The results of the two likelihood analyses are presented in the triangle plot in Fig.~\ref{fig: chains covariance comparison}.
The jackknife and mock-based covariance matrices yield consistent posterior probability distributions, resulting in unbiased estimates of the best-fit parameters.

These results show that the outcomes of the VGCF analyses are only weakly affected by the choice of covariance matrix used in the likelihood analysis, supporting our decision to adopt the matrix estimated with the jackknife method.

The covariant errors estimated through jackknife technique, that relies on a single realization, significantly underestimate the contribution of the sample variance which, instead, is fully included in the errors estimated from the 100 mocks. 
These results therefore also demonstrate that the contribution of the sample variance to the total error budget is negligible. 

\section{Sensitivity to the void size: single mock}
\label{app: rmin single mock}
\begin{figure}
    \centering
    \includegraphics[width=0.9\linewidth]{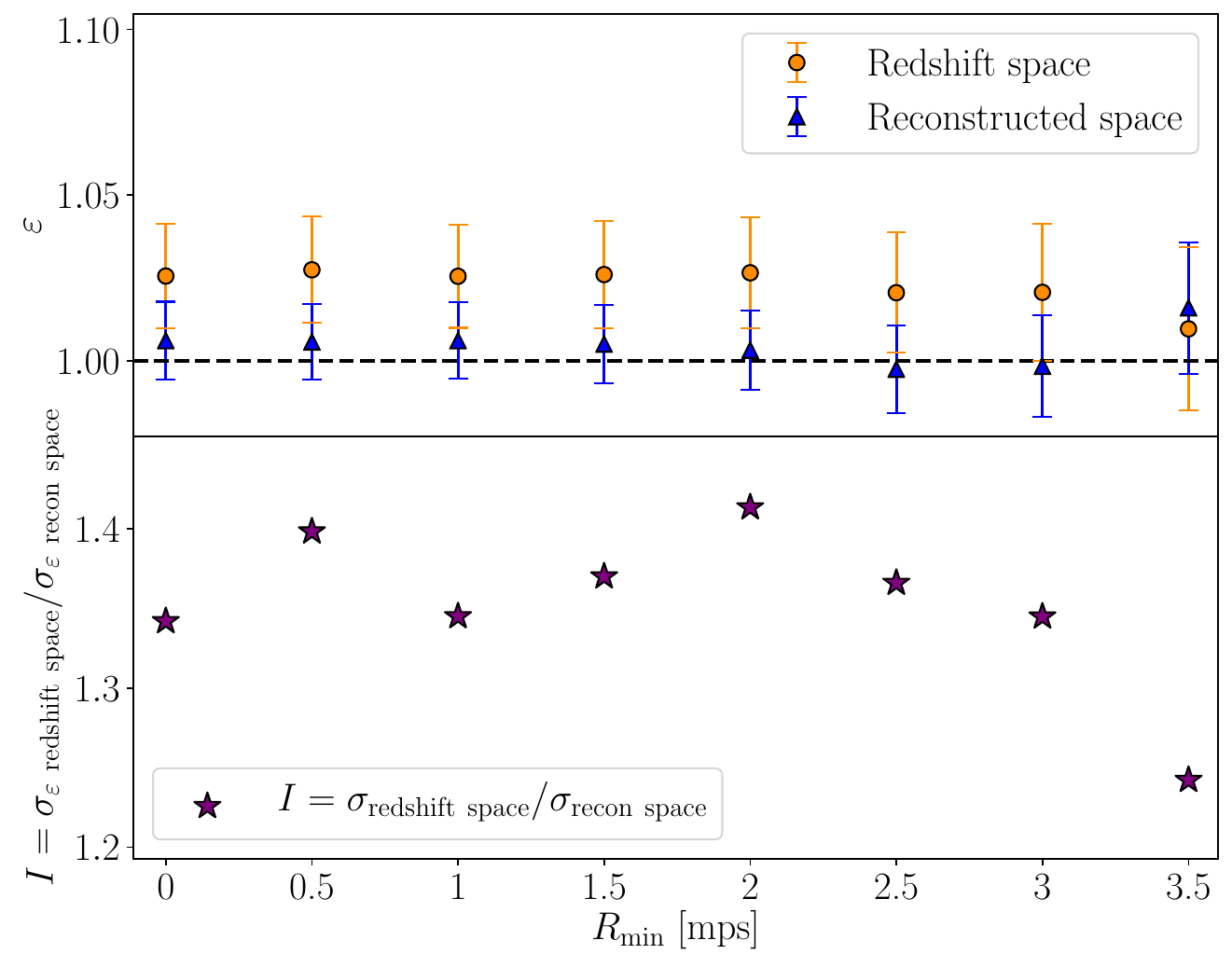}
    \caption{Same as Fig.~\ref{fig: epsilon rmin} but with the analysis performed using only voids extracted from mock 0.}
    \label{fig: epsilon rmin mock 0}
\end{figure}

To illustrate the mock-to-mock scatter discussed in Section~\ref{subsec: void radii}, Fig.~\ref{fig: epsilon rmin mock 0} shows the same $\varepsilon(R_\mathrm{min})$ analysis as in Fig.~\ref{fig: epsilon rmin}, but performed on a single void catalog (mock 0) rather than averaged over 100 mocks. This example highlights that, when small-scale voids are included, the best-fit value of $\varepsilon$ can occasionally deviate from unity in redshift space, whereas reconstructed space consistently yields values compatible with unity. The single-mock analysis therefore provides a concrete illustration of the improved stability and robustness of the reconstruction procedure, which is already quantified in the main text.

\section{Halo bias estimate}
\label{appendix: bias}

To estimate the effective linear bias values of the halos in our mock catalogs, we computed the real-space two-point auto-correlation function in each of the 100 mocks using the Landy–Szalay estimator \citep{landy_szalay}, as implemented in the publicly available package {\fontfamily{cmtt}\selectfont MeasCorr} \citep{farina_2024, guidi_2023}.
We then compared these measurements with the BAO 2PCF model of \citet{ross_2017}, using the mock-based covariance matrix, and inferred the best-fit linear bias value using the {\tt BAOFitter}\footnote{\url{https://gitlab.com/esarpa1/BAOFit}} package.

The cosmological parameters are fixed to those of the mock catalogs, leaving only two free parameters in the {\tt BAOFitter} routine: the linear  bias $b$, and the damping parameter $\Sigma$, which accounts for the suppression of the BAO peak due to nonlinear effects that may also influence the bias measurement.
The best-fit value of $b$ is obtained by minimizing the $\chi^2$ difference between the measurements and the model over the range
 $r=[50,150] \, h^{-1}\text{Mpc}$, yielding $b=1.87 \pm 0.03$.\\

\end{document}